\documentclass[onefignum,onetabnum]{siamonline190516}

\usepackage{algorithmic}
\ifpdf
\DeclareGraphicsExtensions{.eps,.pdf,.png,.jpg}
\else
\DeclareGraphicsExtensions{.eps}
\fi

\usepackage{enumitem}
\setlist[enumerate]{leftmargin=.5in}
\setlist[itemize]{leftmargin=.5in}

\newsiamremark{rem}{Remark}
\newsiamremark{hypothesis}{Hypothesis}
\crefname{hypothesis}{Hypothesis}{Hypotheses}
\newsiamthm{thm}{Theorem}

\headers{Optimal projection for parametric importance sampling in high dimension }{M.El Masri, J.Morio, and F.Simatos}

\title{Optimal projection to improve parametric importance sampling in high dimension}

\author{Maxime El Masri\footnotemark[1] \footnotemark[2]
	\and J\'er\^ome Morio\thanks{ONERA/DTIS, Universit\'e de Toulouse, F-31055 Toulouse, France
		(\email{jerome.morio@onera.fr}, \email{max.elmasri@outlook.fr}).}
	\and Florian Simatos\thanks{ISAE-SUPAERO and Universit\'e de Toulouse, Toulouse, France (\email{florian.simatos@isae.fr}).}}

\usepackage[utf8x,utf8]{inputenc}
\usepackage{lineno,hyperref}
\usepackage{color}
\hypersetup{colorlinks = true, linkcolor = red, citecolor = blue, bookmarksopen = true}
\usepackage{caption}
\usepackage{amsmath, dsfont} 
\usepackage[left=3cm,right=3cm,top=3cm,bottom=3.5cm]{geometry}
\usepackage{stmaryrd}
\usepackage{bbold} 
\usepackage{bm}
\usepackage{epstopdf}
\usepackage[normalem]{ulem}
\usepackage{xcolor}

\usepackage{booktabs}
\usepackage{array}

\usepackage{graphicx}
\usepackage{caption, subcaption}

\usepackage{calc}
\usepackage{ulem}

\usepackage{lmodern}
\usepackage{babel}
\usepackage[all]{xy}
\usepackage{array,multirow}
\usepackage{pgfplots}
\usepackage{tikz,tkz-tab}
\usepackage{ctable, boldline}
\usetikzlibrary{patterns}

\usepackage{multicol}
\usepackage{titletoc}
\usepackage{setspace}

\newcommand{\RR}{\mathds{R}}
\newcommand{\PP}{\mathds{P}}
\newcommand{\EE}{\mathds{E}}
\newcommand{\NO}{\mathcal{N}}
\newcommand{\II}{\mathds{I}}
\newcommand{\xx}{\mathbf{x}}
\newcommand{\XX}{\mathbf{X}}
\newcommand{\mm}{\mathbf{m}}
\newcommand{\dd}{\mathbf{d}}

\newcommand{\Sigmatrois}{{\widehat \Sigma^{\text{}}_\text{opt}}}
\newcommand{\Sigmaquatre}{{\widehat \Sigma^{\text{+d}}_\text{opt}}}
\newcommand{\Sigmacinq}{{\widehat \Sigma^{\text{}}_\text{mean}}}
\newcommand{\Sigmasix}{{\widehat \Sigma^{\text{+d}}_\text{mean}}}

\ifpdf
\hypersetup{
	pdftitle={Optimal projection to improve parametric importance sampling in high dimension},
	pdfauthor={Maxime El Masri, J\'er\^ome Morio, and Florian Simatos}
}
\fi

\begin{document}
	
	\maketitle
	
	\begin{abstract}
		In this paper we propose a dimension-reduction strategy in order to improve the performance of importance sampling in high dimension. The idea is to estimate variance terms in a small number of suitably chosen directions. We first prove that the optimal directions, i.e., the ones that minimize the Kullback--Leibler divergence with the optimal auxiliary density, are the eigenvectors associated to extreme (small or large) eigenvalues of the optimal covariance matrix. We then perform extensive numerical experiments that show that as dimension increases, these directions give estimations which are very close to optimal. Moreover, we show that the estimation remains accurate even when a simple empirical estimator of the covariance matrix is used to estimate these directions. These theoretical and numerical results open the way for different generalizations, in particular the incorporation of such ideas in adaptive importance sampling schemes.
	\end{abstract}
	
	\section{Introduction}
	
	Importance Sampling (IS) is a widely considered stochastic method to estimate integrals of the form $E = \int \phi f$ with a black-box function $\phi$ and a probability density function (pdf) $f$.
	%
	It rests upon the choice of an auxiliary density which, when suitably chosen, can significantly improve the situation compared to the naive Monte Carlo (MC) method~\cite{AgapiouEtAl_ImportanceSamplingIntrinsic_2017, OwenZhou_SafeEffectiveImportance_2000}.
	The theoretical optimal IS density, also called zero-variance density, is defined by $\phi f / E$ when $\phi$ is a positive function. This density is not available in practice as it involves the unknown integral $E$, but a classical strategy consists in searching an optimal approximation in a parametric family of densities. By minimising a ``distance'' with the optimal IS density, such as the Kullback--Leibler divergence, one can find optimal parameters in this family to get an efficient sampling pdf. Adaptive Importance Sampling (AIS) algorithms, such as the Mixture Population Monte Carlo method \cite{CappeEtAl_AdaptiveImportanceSampling_2008}, the Adaptive Multiple Importance Sampling method \cite{CornuetEtAl_AdaptiveMultipleImportance_2012}, or the Cross Entropy method \cite{RubinsteinKroese_CrossentropyMethodUnified_2011}, estimate the optimal parameters adaptively by updating intermediate parameters~\cite{BugalloEtAl_AdaptiveImportanceSampling_2017}.
	
	An intense research activity has made these techniques work very well, but only for moderate dimensions. In high dimension, most of these techniques actually fail to give efficient parameters for two reasons. The first one is the so-called weight degeneracy problem, which is that in high dimension, the weights appearing in the IS densities (which are self-normalized likelihood ratios) degenerate. More precisely, the largest weight takes all the mass, while all other weights are negligible so that the final estimation essentially uses only one sample, see for instance~\cite{BengtssonEtAl_CurseofdimensionalityRevisitedCollapse_2008} for a theoretical analysis in the related context of particle filtering. But even without likelihood ratios, such techniques may fail if they need to estimate high-dimensional parameters such as covariance matrices, whose size increases quadratically in the dimension~\cite{AshurbekovaEtAl_OptimalShrinkageRobust_, LedoitWolf_WellconditionedEstimatorLargedimensional_2004}. 
	The conditions under which importance sampling is applicable in high dimension are notably investigated in a reliability context in \cite{AuBeck_ImportantSamplingHigh_2003}: it is remarked that the optimal covariance matrix should not deviate significantly from the identity matrix. The authors of~\cite{El-LahamEtAl_RecursiveShrinkageCovariance_} tackle the weight degeneracy problem by applying a recursive shrinkage of the covariance matrix, which is constructed iteratively with a weighted sum of the sample covariance estimator and a biased, but more stable, estimator. 
	%
	Concerning the second problem of having to estimate high-dimensional parameters, the idea was recently put forth to reduce the effective dimension by only estimating these parameters (in particular the covariance matrix) in suitable directions~\cite{MasriEtAl_ImprovementCrossentropyMethod_2020, UribeEtAl_CrossentropybasedImportanceSampling_2020}. In this paper we seek to delve deeper into this idea. The main contribution of the present paper is to identify the optimal directions in the fundamental case when the parametric family is Gaussian, and perform numerical simulations in order to understand how they behave in practice. In particular, we propose directions which, in contrast to the recent paper~\cite{UribeEtAl_CrossentropybasedImportanceSampling_2020}, does not require the objective function to be differentiable, and moreover optimizes the Kullback-Leibler distance with the optimal density instead of simply an upper bound on it, as in~\cite{UribeEtAl_CrossentropybasedImportanceSampling_2020}. In Section~\ref{sub:proj} we elaborate in more details on the differences between the two approaches.

	
	
	The paper is organised as follows: in Section~\ref{Sec:IS} we recall the foundations of IS. In Section~\ref{sec:main-result}, we state our main theoretical result and we compare it with the current state-of-the-art. Section~\ref{sec:proof} presents the proof of our theoretical result; Section~\ref{sec:num-results-framework} introduces the numerical framework that we have adopted, and Section~\ref{sec:test-cases} presents the numerical results obtained on five different test cases to assess the efficiency of the directions that we propose. We conclude in Section~\ref{sec:Ccl} with a summary and research perspectives.

	\section{Importance Sampling}\label{Sec:IS}
	
	We consider the problem of estimating the following integral:
	\begin{align*}
	E=\EE_f(\phi(\XX))=\int \phi(\xx)f(\xx)\textrm{d} \xx,
	\end{align*}
	where $\XX$ is a random vector in $\RR^n$ with Gaussian standard pdf $f$, and $\phi: \RR^n\rightarrow\RR_+$ is a real-valued, non-negative function. If one were to relax this Gaussian standard assumption, one would need to look for covariance matrices in a different auxiliary set, see Remark \ref{rem3.2}. The function $\phi$ is considered as a black-box function which is potentially expensive to evaluate, which means the number of calls to $\phi$ should be limited.
	
	
	IS is a widely considered approach to reduce the variance of the classical Monte Carlo estimator of $E$. The idea of IS is to generate a random sample $\XX_1,\ldots,\XX_N$ from an auxiliary density $g$, instead of $f$, and to compute the following estimator: 
	\begin{align}\label{eq:hatE}
	\hat{E}_N=\frac{1}{N}\sum_{i=1}^N \phi(\XX_i)L(\XX_i),
	\end{align}
	with $L=f/g$ the likelihood ratio, or importance weight, and the density $g$, called importance sampling density, is such that $g(\xx)=0$ implies $\phi(\xx) f(\xx)=0$ for every $\xx$ (which makes the product $\phi L$ well-defined). This estimator is consistent and unbiased but its accuracy strongly depends on the choice of the auxiliary density $g$. It is well known that the optimal choice for~$g$ is \cite{bucklew2013introduction}
	\begin{align*}
	g^*(\xx)=\dfrac{\phi(\xx)f(\xx)}{E}, \ \xx\in\RR^n.
	\end{align*}
	Indeed, for this choice we have $\phi L = E$ and so $\hat E_N$ is actually the deterministic estimator~$E$. For this reason, $g^*$ is sometimes called zero-variance density, a terminology that we will adopt here. Of course, $g^*$ is only of theoretical interest as it depends on the unknown integral $E$. However, it gives an idea of good choices for the auxiliary density $g$, and we will seek to approximate~$g^*$ by an auxiliary density that minimizes a distance between~$g^*$ and a given parametric family of densities.
	
	In this paper, the parametric family of densities is the Gaussian family $\{g_{\mm, \Sigma}: \mm \in \RR^n, \Sigma \in \mathcal{S}^+_n\}$, where $g_{\mm, \Sigma}$ denotes the Gaussian density with mean $\mm \in \RR^n$ and covariance matrix $\Sigma \in \mathcal{S}^+_n$ with $\mathcal{S}^+_n \subset \RR^{n \times n}$ the set of symmetric, positive-definite matrices:
	\begin{align*}
	g_{\mm,\Sigma}(\xx)=\dfrac{1}{ \sqrt{ (2\pi)^n \det(\Sigma)} }\exp\left(-\frac{1}{2}(\xx-\mm)^\top\Sigma^{-1}(\xx-\mm)\right), \ \xx \in \RR^n.
	\end{align*}
	Moreover, we will consider the Kullback--Leibler (KL) divergence to measure a ``distance'' between $g^*$ and  $g_{\mm, \Sigma}$. Recall that for two densities $f$ and $h$, with $f$ absolutely continuous with respect to $h$, the KL divergence $D(f,h)$ between $f$ and $h$ is defined by: 
	\begin{align*}
	D(f,h)=\EE_{f}\left[\log \left( \frac{f(\XX)}{h(\XX)} \right) \right] = \int \log \left( \frac{f(\xx)}{h(\xx)} \right)f(\xx) \textrm{d} \xx.
	\end{align*}
	Thus, our goal is to approximate $g^*$ by $g_{\mm^*, \Sigma^*}$ with the optimal mean vector $\mm^*$ and the optimal covariance matrix $\Sigma^*$ given by:
	\begin{equation} \label{eq:argminDkl}
	(\mm^*,\Sigma^*) = \arg\min \left\{ D(g^*,g_{\mm,\Sigma}): \mm \in \RR^n, \Sigma \in \mathcal{S}_n^+ \right\}.
	\end{equation}
	In the Gaussian case of the present setting, it is well-known that $\mm^*$ and $\Sigma^*$ are simply the mean and variance of the zero-variance density~\cite{RubinsteinKroese_CrossentropyMethodUnified_2011,RubinsteinKroese_SimulationMonteCarlo_2017}:
	\begin{align}\label{eq:mstar}
	\mm^*=\EE_{g^*}(\XX) \hspace{0.5cm} \text{ and } \hspace{0.5cm} \Sigma^* = \textrm{Var}_{g^*} \left(\XX\right).
	\end{align}

	\section{Main result and positioning of the paper} \label{sec:main-result}
	
	\subsection{Projecting on a low dimensional subspace} \label{sub:proj}
	
	As $g^*$ is unknown (although, as will be considered below, we can in principle sample from it since it is known up to a multiplicative constant), the optimal parameters $\mm^*$ and $\Sigma^*$ given by~\eqref{eq:mstar} are not directly computable. Therefore, usual estimation schemes start with estimating $\mm^*$ and $\Sigma^*$, say through $\hat \mm^*$ and $\hat \Sigma^*$, respectively, and then use these approximations to estimate $E$ through~\eqref{eq:hatE} with the auxiliary density $g_{\hat \mm^*, \hat \Sigma^*}$. Although the estimation of~$E$ with the auxiliary density $g_{\mm^*, \Sigma^*}$ usually provides very good results, it is well-known that in high dimension, the additional error induced by the estimations of $\mm^*$ and $\Sigma^*$ severely degrades the accuracy of the final estimation \cite{PapaioannouEtAl_ImprovedCrossEntropybased_2019, UribeEtAl_CrossentropybasedImportanceSampling_2020}. 
	The main problem lies in the estimation of $\Sigma^*$ which, in dimension $n$, involves the estimation of a quadratic (in the dimension) number of terms, namely $n(n+1)/2$.
	
	Recently, the idea to overcome this problem by only evaluating variance terms in a small number of influential directions was explored in~\cite{MasriEtAl_ImprovementCrossentropyMethod_2020} and~\cite{UribeEtAl_CrossentropybasedImportanceSampling_2020}. In these two papers, the auxiliary covariance matrix $\Sigma$ is modeled in the form
	\begin{align}\label{eq:Sigmak}
	\Sigma = \sum_{i=1}^k (v_i-1) \dd_i \dd_i^\top + I_n
	\end{align}
	where the $\dd_i$'s are the $k$ orthonormal directions which are deemed influential. It is easy to check that $\Sigma$ is the covariance matrix of the Gaussian vector
	\[ v^{1/2}_1 Y_1 \dd_1 + \cdots + v^{1/2}_k Y_k \dd_k + Y_{k+1} \dd_{k+1} + \cdots + Y_n \dd_n \]
	where the $Y_i$'s are i.i.d.\ standard normal random variables (one-dimensional), and the $n-k$ vectors $(\dd_{k+1}, \ldots, \dd_n)$ complete $(\dd_1, \ldots, \dd_k)$ into an orthonormal basis. In particular, $v_i$ is the variance in the direction of $\dd_i$, i.e., $v_i = \dd_i^\top \Sigma \dd_i$. 
	%
	In~\eqref{eq:Sigmak}, $k$ can be considered as the effective dimension in which variance terms are estimated. In other words, in~\cite{MasriEtAl_ImprovementCrossentropyMethod_2020} and~\cite{UribeEtAl_CrossentropybasedImportanceSampling_2020}, the optimal variance parameter is not sought in $\mathcal{S}^+_n$ as in~\eqref{eq:argminDkl}, but rather in the subset of matrices of the form
	\[ \mathcal{L}_{n,k} = \left\{ \sum_{i=1}^k (\alpha_i-1) \frac{\dd_i \dd_i^\top}{\lVert \dd_i \rVert^2} + I_n: \alpha_1, \ldots, \alpha_k >0 \ \text{ and the $\dd_i$'s are orthogonal} \right\}. \]
	The relevant minimization problem thus becomes
	\begin{equation} \label{eq:argminDkl-k}
	\Sigma^*_k = \arg\min \left\{ D(g^*,g_{\mm^*,\Sigma}): \Sigma \in \mathcal{L}_{n,k} \right\}
	\end{equation}
	instead of~\eqref{eq:argminDkl}, with the effective dimension $k$ being allowed to be adjusted dynamically (see the proof of Theorem~\ref{thm1} for a justification of considering $\mm^*$ instead of optimizing in~$\mm$ also). By restricting the space in which the variance is looked up, one seeks to limit the number of variance terms to be estimated. The idea is that if the directions are suitably chosen, then the improvement of the accuracy due to the smaller error in estimating the variance terms will compensate the fact that we consider less candidates for the covariance matrix.
	
	In~\cite{MasriEtAl_ImprovementCrossentropyMethod_2020}, the authors consider $k = 1$ and $\dd_1 = \mm^* / \lVert \mm^* \rVert$. When $f$ is Gaussian, this choice is motivated by the fact that, due to the light tail of the Gaussian random variable and the reliability context, the variance should vary significantly in the direction of $\mm^*$ and so estimating the variance in this direction can bring information. The method in~\cite{UribeEtAl_CrossentropybasedImportanceSampling_2020} is more involved: $k$ is adjusted dynamically, while the directions $\dd_i$ are the eigenvectors associated to the largest eigenvalues of a certain matrix. They span a low-dimensional subspace called Failure-Informed Subspace, and the authors in~\cite{UribeEtAl_CrossentropybasedImportanceSampling_2020} prove that this choice minimizes an upper bound on the minimal KL divergence. In practice, this algorithm yields very accurate results. However, we will not consider it further in the present paper for two reasons. First, this algorithm is tailored for the reliability case where $\phi = \II_{\{\varphi \geq 0\}}$, \color{black} with a function $\varphi: \RR^n \to \RR$, \color{black} whereas our method is more general and applies to the general problem of estimating an integral (see for instance our test case of Section~\ref{sub:payoff}). Second, the algorithm in~\cite{UribeEtAl_CrossentropybasedImportanceSampling_2020} requires the evaluation of the gradient of the function $\varphi$. However, this gradient is not always known and can be expensive to evaluate in high dimension; in some cases, the function $\varphi$ is even not differentiable, as will be the case in our numerical example in Section~\ref{sub:portfolio}.  
	In contrast, our method makes no assumption on the form or smoothness of~$\phi$: it does not need to assume that it is of the form $\II_{\{\varphi \geq 0\}}$, or to assume that $\nabla \varphi$ is tractable. For completeness, whenever  the algorithm of~\cite{UribeEtAl_CrossentropybasedImportanceSampling_2020} was applicable and computing the gradient of $\varphi$ did not require any additional simulation budget, we have run it on the test cases considered here and found that it outperformed our algorithm (see Remark~\ref{rem:FIS} below). In more realistic settings, computing $\nabla \varphi$ would likely increase the simulation budget, and it would be interesting to compare the two algorithms in more details to understand when this extra computation cost is worthwhile. We reserve such a question for future research and will not consider the algorithm of~\cite{UribeEtAl_CrossentropybasedImportanceSampling_2020} further, as our aim in this paper is to establish benchmark results for a general algorithm which works for any function $\phi$. \color{black}

	\subsection{Main result of the paper} \label{sub:main-result+positioning}
	
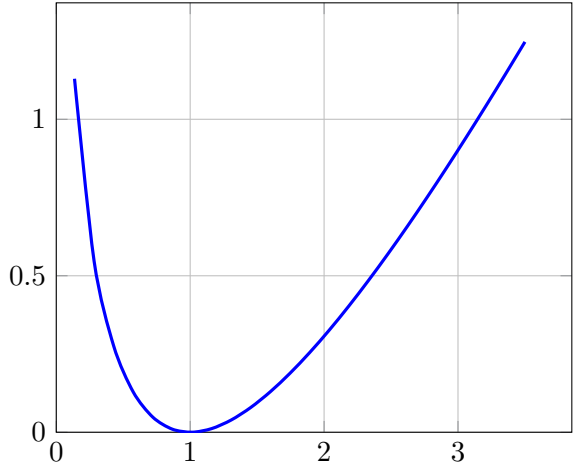
\begin{figure}[t]
		\centering
		\begin{tikzpicture}
		\pgfplotsset{work well/.style={mark=none, dashed, very thick},
			work not/.style={mark=none, solid, very thick}}
		\begin{axis}[grid,
		xmin=0, ymin=0, ytick={0,0.5,1}
		]
		\addplot [blue, smooth, domain=-.01:3.5, very thick] {-ln(x)+x-1};
		
		\end{axis}
		\end{tikzpicture}
		\caption{Plot of the function $\ell(x) = -\log x + x - 1$ given by~\eqref{eq:l}.}
		\label{fig:l}
	\end{figure}

	The main result of the present paper is to actually compute the exact value for $\Sigma^*_k$ in~\eqref{eq:argminDkl-k}, which therefore paves the way for efficient high-dimensional estimation schemes. The statement of our result involves the following function~$\ell$, which is represented in Figure~\ref{fig:l}:
	\begin{align}\label{eq:l}
	\ell: x \in (0,\infty) \mapsto -\log(x) + x - 1.
	\end{align}
	In the following, $(\lambda, \dd) \in \RR \times \RR^n$ is an eigenpair of a matrix $A$ if $A\dd = \lambda \dd$ and $\lVert \dd \rVert = 1$. A diagonalizable matrix has $n$ distinct eigenpairs, say $((\lambda^*_i, \dd^*_i), i = 1, \ldots, n)$, and we say that these eigenpairs are ranked in decreasing $\ell$-order if $\ell(\lambda^*_1) \geq \cdots \geq \ell(\lambda^*_n)$.
	
	\
	
	\begin{center}\fbox{\begin{minipage}{0.95\textwidth}
				\begin{thm}\label{thm1}
					Let $(\lambda^*_i, \dd^*_i)$ be the eigenpairs of $\Sigma^*$ ranked in decreasing $\ell$-order. Then for $1 \leq k \leq n$, the solution $\Sigma^*_k$ to~\eqref{eq:argminDkl-k} is given by
					\begin{equation}
					\Sigma^*_k = I_n + \sum_{i=1}^k \left( \lambda^*_i - 1 \right) \dd^*_i (\dd^*_i)^\top. \label{eq:Sigma*k}
					\end{equation}
					%
				\end{thm}
	\end{minipage}}\end{center}
	
	\medskip
	
	For $k = 1$ for instance, the shape of the function $\ell$ depicted in Figure~\ref{fig:l} implies that $\Sigma^*_1 = I_n + (\lambda^*-1) \dd^* (\dd^*)^\top$ with $(\lambda^*, \dd^*)$ the eigenpair of $\Sigma^*$ with $\lambda^*$ either the largest or the smallest eigenvalue of $\Sigma^*$, depending on which one maximizes $\ell$.

	
	This theoretical result therefore suggests to reduce dimension by estimating eigenpairs of~$\Sigma^*$, rank them in decreasing $\ell$-order and then use the $k$ first eigenpairs $(({\hat \lambda}^*_i, {\hat \dd}^*_i), i = 1, \ldots, k)$ to build the covariance matrix $\hat \Sigma^*_k = \sum_{i=1}^k ({\hat \lambda}^*_i-1) {\hat \dd}^*_i ({{\hat \dd}^*}_i)^\top + I_n$ and the corresponding auxiliary density. This scheme is summarized in Algorithm~\ref{algo:ISprojopt}. The effective dimension~$k$ is obtained by Algorithm~\ref{algo:choicek}, see Section~\ref{sub:choicek} below.

	\begin{algorithm}[h]
	\begin{algorithmic}
\STATE \textbf{Data: } Sample sizes $N$ and $M$
\STATE \textbf{Result: } Estimation $\hat{E}_N$ of integral $E$
		\STATE - Generate a sample $\XX_1,\ldots,\XX_M$  on $\RR^n$ independently according to $g^*$ 
		
		\STATE - Estimate $\hat{\mm}^*$ and $\hat{\Sigma}^*$ defined in~\eqref{eq:hatm} and~\eqref{eq:hatSigma} with this sample\;
		\STATE - Compute the eigenpairs $(\hat \lambda^*_i, \hat \dd^*_i)$ of $\hat \Sigma^*$ ranked in decreasing $\ell$-order\;
		\STATE - Compute the matrix $\hat \Sigma^*_k = \sum_{i=1}^k ({\hat \lambda}^*_i-1) {\hat \dd}^*_i ({{\hat \dd}^*}_i)^\top + I_n$ with $k$ obtained by applying Algorithm~\ref{algo:choicek} with input $({\hat \lambda}^*_1, \ldots, {\hat \lambda}^*_n)$\;
		\STATE - Generate a new sample $\XX_1,\ldots,\XX_N$ independently from $g' = g_{\hat \mm^*,\hat \Sigma^*_k}$\;
		\STATE - Return $\displaystyle \hat{E}_N=\frac{1}{N}\underset{i=1}{\overset{N}{\sum}} \phi(\XX_i)\frac{f(\XX_i)}{g'(\XX_i)}$.
	\end{algorithmic}
		\caption{Algorithm suggested by Theorem~\ref{thm1}.}
		\label{algo:ISprojopt}
	\end{algorithm}
	%

    \begin{rem} \label{rem3.2}
    The value $1$ plays a particular role in Theorem~\ref{thm1}, in that, as $\ell$ is minimized in $1$, eigenvectors with eigenvalues $1$ will only be selected once all other eigenvalues will have been picked: in other words, if $\lambda^*_i = 1$ then $\lambda^*_j = 1$ for all $j \geq i$. The reason why $1$ plays this special role is due to the form of the covariance matrix that we impose. More precisely, looking for covariance matrices in the set $\mathcal{L}_{n,k}$ amounts to looking for covariance matrices which, once diagonalized, have one's on the diagonal except possibly for $k$ values (the $\alpha_i$'s). As $k$ will be small, typically $k = 1$ or $2$, this amounts to looking for covariance matrices which are perturbation of the identity. This is particularly relevant as we assume $f$ is a standard Gaussian density. What Theorem~\ref{thm1} tells is that, when trying to approximate $\Sigma^*$ by such matrices, we should first consider eigenvectors with eigenvalues as different as possible from $1$, the ``distance'' to $1$ being measured by $\ell$. If one was imposing a different form on $\Sigma^*_k$ (which can be interesting if the distribution $f$ is not standardized), then a different result would arise. For instance, if one was looking for matrices where the ``default'' choice would be some $\lambda > 0$ for the diagonal entries that are not estimated, i.e., a matrix of the form $\sum_i (\alpha_i-\lambda) \dd_i \dd_i^\top + \lambda I_n$, then eigenpairs would be ranked according to the function $\ell(\cdot/\lambda)$, meaning that one would look for eigenvectors associated to eigenvalues as different as possible from $\lambda$.
	\end{rem}

	\color{black}

As mentioned above, we assume in the first step of Algorithm~\ref{algo:ISprojopt} that we can sample according to $g^*$. Since $g^*$ is known up to a multiplicative constant, this is a reasonable assumption as classical techniques such as importance sampling with self-normalized weights or Markov Chain Monte--Carlo (MCMC) can be applied in this case (see for instance \cite{ChanKroese_ImprovedCrossentropyMethod_2012, GraceEtAl_AutomatedStateDependentImportance_2014}). In this paper, we choose to apply a basic rejection method that yields perfect independent samples from $g^*$, possibly at the price of a high computational cost. As the primary goal of this paper is to understand whether the $\dd^*_i$'s are indeed good projection directions, this computational cost to generate from $g^*$ is not relevant for us and therefore not taken into account. Possible improvements to relax this assumption are discussed in the conclusion of the paper.

	\subsection{Choice of the number of dimensions $k$}\label{sub:choicek}
	
	The choice of the effective dimension $k$, i.e., the number of projection directions considered, is important. If it is close to $n$, then the matrix $\hat \Sigma^*_k$ will be close to $\hat \Sigma^*$ which is the situation we want to avoid in the first place. On the other hand, setting $k=1$ in all cases may be too simple and lead to suboptimal results. In practice however, this is often a good choice. In order to adapt $k$ dynamically, we consider a simple method based on the value of the KL divergence. 
	Given the eigenvalues $\lambda_1, \ldots, \lambda_n$ ranked in decreasing $\ell$-order, we look for the maximal gap in the sequence $(\ell(\lambda_1), \ldots, \ell(\lambda_n))$. This allows to choose $k$ such that $\sum_{i=1}^k \ell(\lambda_i)$ is close to $\sum_{i=1}^n \ell(\lambda_i)$ which is equal, up to an additive constant, to the minimal KL divergence (see~\eqref{eq:D'-ell} below). The precise method is described in Algorithm~\ref{algo:choicek}. 
	\begin{algorithm}[h]
	\begin{algorithmic}
\STATE \textbf{Data: } Sequence of positive numbers $\lambda_1, \ldots, \lambda_n$ in decreasing $\ell$-order
\STATE \textbf{Result: } Number of selected dimensions $k$
	\STATE	- Compute the increments $\delta_i = \ell(\lambda_{i+1}) - \ell(\lambda_i)$ for $i=1\ldots n-1$
	\STATE	- Return $k=\arg\max \delta_i$, the index of the maximum of the differences.
	\end{algorithmic}
		\caption{Choice of the number of dimensions}
		\label{algo:choicek}
	\end{algorithm}

	\section{Proof of Theorem~\ref{thm1}} \label{sec:proof}
	
	For any $\mm \in \RR^n$ and $\Sigma \in \mathcal{S}^+_n$, we have by definition
	\[ D(g^*, g_{\mm, \Sigma}) = \EE_{g^*} \left( \log \left( \frac{g^*(\XX)}{g_{\mm, \Sigma}(\XX)} \right) \right). \]
	Plugging in $g_{\mm, \Sigma}(\xx) = (2 \pi)^{-n/2} (\det \Sigma)^{-1/2} e^{-(\xx-\mm)^\top \Sigma^{-1} (\xx - \mm)}$ we obtain
	\[ D(g^*, g_{\mm, \Sigma}) = \frac{1}{2} \log \det \Sigma + \frac{1}{2} \EE_{g^*} \left( (\XX - \mm)^\top \Sigma^{-1} (\XX - \mm) \right) + C \]
	with $C$ a term independent of $\mm$ and $\Sigma$, therefore treated as a constant when trying to optimize in these variables.
	The gradient in $\mm$ is equal to $\Sigma^{-1} \EE_{g^*}(\XX - \mm)$. This implies that the function is necessarily minimized in $\mm = \mm^*$, which will be assumed for the rest of the proof (and also justifies setting $\mm = \mm^*$ in~\eqref{eq:argminDkl-k}). 
	In the rest of the proof (and of the paper) we will focus on the remaining terms and define
	\[ D'(\Sigma) = \log\det\Sigma + \EE_{g^*}\left((\XX-\mm^*)^\top\Sigma^{-1}(\XX-\mm^*)\right). \]
	Using the identity $a^\top b = \textrm{tr}(a b^\top)$ for any two column vectors, we see that
	\[ \EE_{g^*} \left[ (\XX-\mm^*)^\top \Sigma^{-1} (\XX-\mm^*) \right] = \EE_{g^*} \left[ \textrm{tr} \left( (\XX-\mm^*) (\XX-\mm^*)^\top \Sigma^{-1} \right) \right]. \]
	Since the expectation and trace operators commute, using the linearity of the expectation and the fact that $\Sigma^* = \textrm{Var}_{g^*}(\XX)$, we finally obtain
	\begin{equation} \label{eq:D'}
	D'(\Sigma) = \log\det\Sigma + \textrm{tr}\left( \Sigma^* \Sigma^{-1} \right).
	\end{equation}
	The goal is to minimize this quantity over $\Sigma \in \mathcal{L}_{n,k}$. In the rest of the proof, we consider $\mathbf{v} = (v_1, \ldots, v_k) \in (0,\infty)^k$ and $\dd = (\dd_1, \ldots \dd_k)$ orthogonal, and we are interested in minimizing $D'(\Sigma)$ with $\Sigma = \sum_{i=1}^k (v_i-1) \dd_i \dd_i^\top / \lVert \dd_i \rVert^2 + I_n$. Thus, in the sequel, we see $D'$ as a function of $\mathbf{v}$ and $\dd$.
	
	
	\noindent \textit{First step: computation of $D'(\Sigma)$.} The goal of this first step is to prove that
	\begin{align}\label{eq:D'-2}
	D'(\Sigma) = D' \left( \mathbf{v}, \dd \right) = \sum_{i=1}^k \left[ \log(v_i) + \left( \frac{1}{v_i}-1 \right) \psi(\dd_i) \right] + \textrm{tr}(\Sigma^*)
	\end{align}
	where here and in the sequel, $\Sigma = \sum_{i=1}^k (v_i-1) \dd_i \dd_i^\top / \lVert \dd_i \rVert^2 + I_n$ and $\psi(\xx) = \xx^\top \Sigma^* \xx / \lVert \xx \rVert^2$. To do so, note that $\Sigma = Q \Delta Q^\top$, with $\Delta = \textrm{diag}(v_1, \ldots, v_k, 1, \ldots, 1)$ a diagonal matrix and~$Q$ an orthogonal matrix with $\dd_i / \lVert \dd_i \rVert$ as $k$ first columns. Indeed, we have $\Sigma \dd_i = v_i \dd_i$, so that $\dd_i$ is an eigenvector with eigenvalue $v_i$. Moreover, for all $\dd_\bot$ in the orthogonal space of $\textrm{span}(\dd)$, we have $\Sigma \dd_\bot = \dd_\bot$, so that $1$ is eigenvalue of multiplicity $n-k$ (or more if some of the $v_i$'s are equal to one).
	
	It follows from this decomposition that $\det(\Sigma) = v_1 \cdots v_k$. Moreover, we have
	\[ \textrm{tr} \left( \Sigma^*\Sigma^{-1} \right) = \textrm{tr} \left( \Sigma^* Q\Delta^{-1}Q^\top \right) = \textrm{tr} \left( \Delta^{-1}Q^\top\Sigma^* Q \right). \]
	Since the first columns of $Q$ are $\dd_i / \lVert \dd_i \rVert$, the first diagonal coefficients of $Q^\top \Sigma^* Q$ are $\psi(\dd_i)$. Thus, if $\dd_{k+1}, \ldots, \dd_n$ complete the $\dd_i / \lVert \dd_i \rVert$ into an orthonormal basis, we have
	\[ \textrm{tr} \left( \Delta^{-1}Q^\top\Sigma^* Q \right) = \sum_{i=1}^k \frac{1}{v_i} \psi(\dd_i) + \sum_{i=k+1}^n \psi(\dd_i) = \sum_{i=1}^k \left( \frac{1}{v_i}-1 \right) \psi(\dd_i) + \sum_{i=1}^n \psi(\dd_i). \]
	Since the last sum $\sum_{i=1}^n \psi(\dd_i)$ is equal to $\textrm{tr}(Q^\top \Sigma^* Q) = \textrm{tr}(\Sigma^*)$, gathering the previous equalities leads to~\eqref{eq:D'-2}.
	\\
	
	\noindent \textit{Second step: minimization.} Starting from~\eqref{eq:D'-2}, we see that the derivative in $v_i$ is given by
	\[ \dfrac{\partial D'}{\partial v_i}(\mathbf{v}, \dd) = \dfrac{1}{v_i} - \frac{1}{v_i^2} \psi(\dd_i) = \frac{1}{v_i^2} \left( v_i - \psi(\dd_i) \right). \]
	Thus, for fixed $\dd$, $D'$ is decreasing in $v_i$ for $v_i < \psi(\dd_i)$ and then increasing for $v_i > \psi(\dd_i)$, which shows that, for fixed $\dd$, $D'$ is minimized for $v_i = \psi(\dd_i)$. For this value, say $\textbf{v}^* = (\psi(\dd_1), \ldots, \psi(\dd_k))$, we have
	\begin{equation} \label{eq:D'-ell}
	D'(\mathbf{v}^*, \dd) = \sum_{i=1}^k \left[ \log(\psi(\dd_i)) + 1 - \psi(\dd_i) \right] + \textrm{tr}(\Sigma^*) = \textrm{tr}(\Sigma^*)-\sum_{i=1}^k \ell(\psi(\dd_i)) .
	\end{equation}
	Since $\ell$ is decreasing and then increasing, it is clear from this expression that in order to minimize $D'$, one must choose the $\dd_i$'s in order to either maximize or minimize $\psi$, whichever maximizes $\ell$. Since the variational characterization of eigenvalues shows that eigenvectors precisely solve this problem, we get the desired result.

	\section{Framework for the numerical results} \label{sec:num-results-framework}
	
	\subsection{General framework}

	The objective of the numerical simulations is to evaluate the impact of the choice of the covariance matrix on the estimation accuracy of a high dimensional integral $E$. We compare in this section the estimation results for different choices of the auxiliary covariance matrix when the IS auxiliary density is Gaussian. To extend this comparison, we also compute the results when the IS auxiliary density is chosen with the von Mises–Fisher– Nakagami (vMFN) model recently proposed in \cite{PapaioannouEtAl_ImprovedCrossEntropybased_2019} for high dimensional probability estimation.

In the following section we test these different models of auxiliary densities on five test cases, where, except for the third one (Section~\ref{sub:banana}), $f$ is a standard Gaussian density. This choice is not a theoretical limitation as we can in principle always come back to this case by transforming the vector~$\XX$ with isoprobabilistic transformations (see for instance~\cite{HohenbichlerRackwitz_NonNormalDependentVectors_1981, LiuDerKiureghian_MultivariateDistributionModels_1986}).

The precise numerical framework that we will consider to assess the efficiency of the different auxiliary models is as follows. We assume first that $M$ i.i.d.\ random samples $\XX_1,\ldots,\XX_M$ distributed from $g^*$ are available from rejection sampling. From these samples, the parameters of the Gaussian and of the vMFN auxiliary density are computed to get an auxiliary density $g'$. Finally, $N$ samples are generated from $g'$ to provide an estimation of $E$ with IS. This procedure is summarized by the following stages: 
	\begin{enumerate}
	    \item Generate a sample $\XX_1,\ldots,\XX_M$ independently according to $g^*$;
	   \item From $\XX_1,\ldots,\XX_M$, compute the parameters of the auxiliary parametric density $g'$;
 \item  Generate a new sample $\XX_1,\ldots,\XX_N$ independently from $g'$;
		\item   Estimate $E$ with $\hat{E}_N=\frac{1}{N}\underset{i=1}{\overset{N}{\sum}} \phi(\XX_i)\frac{f(\XX_i)}{g'(\XX_i)}$.
	\end{enumerate}
	The number of samples $M$ and $N$ are respectively set to $M=500$ and $N=2000$. This procedure is then repeated $50$ times to provide a mean estimation $\hat E$ of $E$. In the result tables, for each auxiliary density $g'$ we report the corresponding value for the relative error $\hat E/ E-1$ and the coefficient of variation of the $50$ iterations (the empirical standard deviation divided by $E$). As was established in the proof of Theorem~\ref{thm1}, the KL divergence is, up to an additive constant, equal to $D'(\Sigma) = \log \textrm{det} \Sigma + \textrm{tr}(\Sigma^* \Sigma^{-1})$ which we will refer to as partial KL divergence. In the result tables, we also report thus the mean value of $D'(\Sigma)$ to analyse the relevancy of the auxiliary density $g_{\hat \mm^*, \Sigma}$ for these six choices of covariance matrix $\Sigma$. The next sections specify the different parameters of $g'$ for the Gaussian model and for the vMFN model we have considered in the simulations.

	\subsubsection{Choice of the auxiliary density $g'$ for the Gaussian model}\label{def_cov}
		The goal is to get benchmark results to assess whether one can improve estimations of Gaussian IS auxiliary density by projecting the covariance matrix~$\Sigma^*$ in the proposed directions $\dd^*_i$. The algorithm that we study here (Algorithms~\ref{algo:ISprojopt}+\ref{algo:choicek}) aims more precisely at understanding whether:
	\begin{itemize}
		\item projecting can improve the situation with respect to the empirical covariance matrix;
		\item the $\dd^*_i$'s are good candidates, in particular compared to the choice $\mm^*$ suggested in~\cite{MasriEtAl_ImprovementCrossentropyMethod_2020};
		\item what is the impact in making errors in estimating the eigenpairs $(\lambda^*_i, \dd^*_i)$.
	\end{itemize}
	Let us define the estimate  $\hat \mm^*$ of $\mm^*$	from the $M$ i.i.d.\ random samples $\XX_1,\ldots,\XX_M$ distributed from $g^*$ with
	\begin{equation} \label{eq:hatm}
	\hat{\mm}^* = \frac{1}{M}\sum_{i=1}^M \XX_i.
	\end{equation}
In our numerical test cases, we will compare six different choices of Gaussian auxiliary distributions $g'$ with mean $\hat{\mm}^*$ and the following covariance matrices (see Table~\ref{tab:sigma}):
	\begin{enumerate}
		\item $\Sigma^*$: the optimal covariance matrix given by~\eqref{eq:mstar};
		\item $\hat{\Sigma}^*$: the empirical estimation of $\Sigma^*$ given by
	\begin{equation}\label{eq:hatSigma}
	\hat \Sigma^* = \frac{1}{M}\sum_{i=1}^M (\XX_i-\hat{\mm}^*)(\XX_i-\hat{\mm}^*)^\top.
	\end{equation}
	\end{enumerate}
	The four other covariance matrices considered in the numerical simulations are of the form 
	 $\sum_{i=1}^k (v_i-1) \dd_i \dd^\top_i + I_n$ where $v_i$ is the variance of $\hat \Sigma^*$ in the direction $\dd_i$, $v_i = \dd_i^\top \hat \Sigma^* \dd_i$. The considered choice of $k$ and $\dd_i$ gives the following covariance matrices:   
	\begin{enumerate}[resume]
		\item $\Sigmatrois$ is obtained by choosing $\dd_i = \dd^*_i$ of Theorem~\ref{thm1}, which is supposed to be perfectly known from $\Sigma^*$ and $k$ is computed with Algorithm \ref{algo:choicek};
		\item $\Sigmaquatre$ is obtained by choosing $\dd_i = {\hat \dd}^*_i$ the $i$-th eigenvector of $\hat \Sigma^*$ (in $\ell$-order), which is an estimation of $\dd^*_i$, and $k$ is computed with Algorithm \ref{algo:choicek};
		\item $\Sigmacinq$ is obtained by choosing $k = 1$ and $\dd_1 = \mm^* / \lVert \mm^* \rVert$;
		\item $\Sigmasix$ is obtained by choosing $k = 1$ and $\dd_1 = {\hat \mm}^* / \lVert {\hat \mm}^* \rVert$, where $\hat \mm^*$ given by~\eqref{eq:hatm}.
	\end{enumerate}

The matrices $\Sigmatrois$ and $\Sigmacinq$ use the estimation $\hat \Sigma^*$ but the actual directions $\dd^*_i$ or $\mm^*$, while the matrices $\Sigmaquatre$ and $\Sigmasix$ involve an additional estimation of the directions. By definition, $\Sigma^*$ will give optimal results, while results for $\hat \Sigma^*$ will deteriorate as the dimension increases, which is the well-known behavior which we try to improve. Moreover, for $\Sigmacinq$ and $ \Sigmatrois$, the projection directions, if not known analytically, are obtained by a brute force Monte Carlo scheme with a very high simulation budget. Finally, we emphasize that Algorithm~\ref{algo:ISprojopt} corresponds to estimating and projecting on the $\dd^*_i$'s, and so the matrix $\hat \Sigma^*_k$ of Algorithm~\ref{algo:ISprojopt} is equal to the matrix $\Sigmaquatre$, i.e., $\hat \Sigma^*_k = \Sigmaquatre$.

 \newlength{\tmpln}
 \settowidth{\tmpln}{Choice for the projection directions}
 
 \begin{table}[]
     \centering
			\begin{tabular}{|l|c|c|c|c|c|c|}
				\hline
			 \parbox{10mm}{\vspace{2pt} \color{white} $\Sigma^*$  \vspace{2pt}} & $\Sigma^*$ & $\hat{\Sigma}^*$ & $\Sigmatrois$ & $\Sigmacinq$ & $\Sigmaquatre$ & $\Sigmasix$\\
				\hline
			 Initial covariance matrix & $\Sigma^*$ & \multicolumn{5}{c|}{$\hat \Sigma^*$} \\ \hline
 			 Projection directions: exact or estimated & \multicolumn{2}{c|}{---} & \multicolumn{2}{c|}{Exact} & \multicolumn{2}{c|}{Estimated} \\ \hline
			 Choice for the projection directions & \multicolumn{2}{c|}{None} & Opt & Mean & Opt & Mean \\\hline
			\end{tabular}
     \caption{
     Presentation of the six covariance matrices considered in the numerical examples. Except $\Sigma^*$, the five other matrices involve one or two estimations: $\hat \Sigma^*$ is the empirical estimation of $\Sigma^*$ given by~\eqref{eq:hatSigma}. The four others are obtained by projecting $\hat \Sigma^*$ on: (i)~the optimal directions $\dd^*_i$ for $\Sigmatrois$; (ii)~estimations $\hat \dd^*_i$ of the optimal directions $\dd^*_i$ for $\Sigmaquatre$; (iii)~$\mm^*$ for $\Sigmacinq$; and (iv)~the estimation $\hat \mm^*$~\eqref{eq:mstar} of $\mm^*$ for $\Sigmasix$. The subscript therefore indicates the choice for the projection direction, while the superscript +d indicates whether these directions are estimated or not.}
     \label{tab:sigma}
 \end{table}

	\subsubsection{Choice of the auxiliary density $g'$ for the von Mises–Fisher–Nakagami model}
	Von Mises–Fisher–Nakagami (vMFN) distributions were proposed in~\cite{PapaioannouEtAl_ImprovedCrossEntropybased_2019} as an alternative to the Gaussian parametric family to perform IS for high dimensional probability estimation. A random vector $\XX$ drawn according to the vMFN distribution can be written as $\XX=R {\bf A}$ where ${\bf A}=\frac{\XX}{\lVert\XX\rVert}$ is a unit random vector following the von Mises--Fisher distribution, and $R=\lVert\XX\rVert$ is a positive random variable with a Nakagami distribution; further, $R$ and $\bf A$ are independent. The vMFN pdf can be written as
	\begin{equation}
	    g_\text{vMFN}({\bf x})= g_\text{N}(\lVert{\bf x}\rVert, p, \omega) \times g_\text{vMF} \left( \frac{{\bf x}}{\lVert{\bf x}\rVert}, {\bm\mu}, \kappa \right).
	    \label{vMFN}
	\end{equation}
The density $g_\text{N}(\lVert {\bf x}\rVert, p, \omega)$ is the Nakagami distribution with shape parameter $p \geq 0.5$ and a spread parameter $\omega>0$ defined by
\[ g_\text{N}(\lVert {\bf x}\rVert, p, \omega) = \frac{2 p^p}{\Gamma(p) \omega^p} \lVert {\bf x}\rVert^{2p-1} \exp\left( - \frac{p}{\omega}\lVert {\bf x}\rVert^2\right) \]
and the density $g_\text{vMF}(\frac{{\bf x}}{\lVert{\bf x}\rVert}, {\bm\mu}, \kappa)$ is the von Mises--Fisher distribution, given by
\[ g_\text{vMF} \left( \frac{{\bf x}}{\lVert{\bf x}\rVert}, {\bm\mu}, \kappa \right) = C_n(\kappa) \exp\left(\kappa {\bm\mu}^T \frac{{\bf x}}{\lVert{\bf x}\rvert\rvert} \right),\]
where $C_n(\kappa) $ is a normalizing constant, $\bm \mu$ is a mean direction $\bm\mu$ (with $\lvert\lvert\bm\mu\rvert\rvert=1$) and $\kappa > 0$ is a concentration parameter $\kappa>0$.

Choosing a vMFN distribution therefore amounts to choosing the parameters $p, \omega, {\bm \mu},$ and $\kappa$. There are therefore $n+3$ parameters to estimate, which is a significant reduction compared to the $\frac{n(n+3)}{2}$ required parameters of the Gaussian model with full covariance matrix.

Following~\cite{PapaioannouEtAl_ImprovedCrossEntropybased_2019}, given a sample $\XX_1,\ldots,\XX_M$ distributed from $g^*$, the parameters $\omega$, $p$, $\bm \mu$ and $\kappa$ are set in the following way in order to define $g'$:
\[ \widehat{\omega}=\frac{1}{M}\sum_{i=1}^M \lVert\XX_i\rVert^2 \ \text{ and } \ \widehat{p}=\frac{\widehat{\omega}^2}{\widehat{\tau}-\widehat{\omega}^2} \text{ with } \widehat{\tau}=\frac{1}{M}\sum_{i=1}^M \lVert\XX_i\rVert^4 \]
and
\[ \widehat{\bm\mu}=\frac{\sum_{i=1}^M \frac{\XX_i}{\lvert\lvert\XX_i\rvert\rvert}}{\lvert\lvert\sum_{i=1}^M \frac{\XX_i}{\lvert\lvert\XX_i\rvert\rvert} \rvert\rvert} \ \text{ and } \  \widehat{\kappa}=\dfrac{n\widehat{\chi}-\widehat{\chi}^3}{1-\widehat{\chi}^2} \text{ with } \widehat{\chi} = \min \left( \left \lVert \frac{1}{M}\sum_{i=1}^M \frac{\XX_i}{\lVert \XX_i \rVert} \right \rVert, 0.95 \right). \]

 	\color{black}

 	\section{Numerical results on five test cases} \label{sec:test-cases}
 	The proposed numerical framework is applied on three examples that are often considered to assess the performance of importance sampling algorithms and also two test cases  from the area of financial mathematics. Extended simulation results are given in the supplementary material associated with this article. 	\color{black}  
 	
	\subsection{Test case 1: one-dimensional optimal projection}\label{sub:sum}

	We consider a test case where all computations can be made exactly. This is a classical example of rare event probability estimation, often used to test the robustness of a method in high dimension. It is given by $\phi(\xx)=\II_{\{\varphi(\xx)\geq 0\}}$ with $\varphi$ the following affine function:
	\begin{align}\label{eq:sum}
	\varphi: \xx=(x_1,\ldots,x_n)\in\RR^n \mapsto\underset{j=1}{\overset{n}{\sum}} x_j-3\sqrt{n}.
	\end{align}
	The quantity of interest $E$ is defined as $E=\int_{\mathbb{R}^n} \phi(\xx) f(\xx) \textrm{d}\xx = \PP_f(\varphi(\XX)\geq 0)\simeq 1.35\cdot 10^{-3}$ for all $n$ where the density $f$ is the standard $n$-dimensional Gaussian distribution. \color{black} Here, the zero-variance density is $g^*(\xx)=\dfrac{f(\xx)\II_{\{\varphi(\xx)\geq 0\}}}{E}$, and the optimal parameters~$\mm^*$ and~$\Sigma^*$ in \eqref{eq:mstar} can be computed exactly, namely $\mm^* = \alpha \textbf{1}$ with $\alpha = e^{-9/2}/(E(2\pi)^{1/2})$ and $\textbf{1} = \frac{1}{\sqrt n} (1,\ldots,1) \in \RR^n$ the normalized constant vector, and $\Sigma^* =(v-1) \mathbf{1} \mathbf{1}^\top + I_n$ with $v=3\alpha-\alpha^2+1$. 

	\subsubsection{Evolution of the partial KL divergence and spectrum}
	
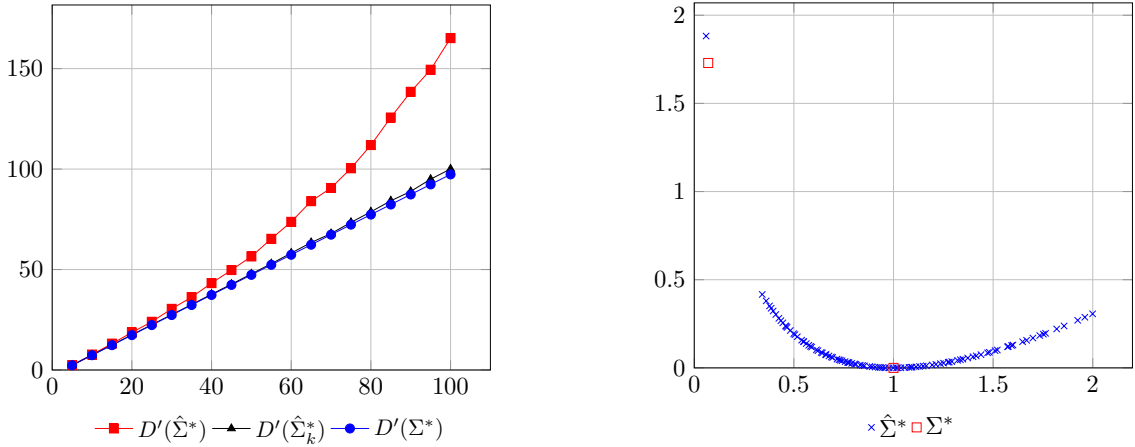
\begin{figure}[t]
		\begin{subfigure}[t]{.45\textwidth}
			\centering
			\begin{tikzpicture}[scale=.85]
			\begin{axis}[
			xlabel={dimension $n$}, xmin=0, ymin=0,
			ylabel={},
			legend style={at={(.5,-.1)}, anchor=north, draw=none, font=\small}, grid, legend columns = 3
			]
			\addplot [ red, mark=square*, mark options={fill=red}] table {Data2/Dkl1.txt};
			\addlegendentry{$D'(\hat \Sigma^*)$}
			
			\addplot [ black, mark=triangle*, mark options={fill=black}] table {Data2/Dklproj.txt};
			\addlegendentry{$D'(\hat \Sigma^*_k)$}
			\addplot [ blue, mark=*, mark options={fill=blue}] table {Data2/Dklstar.txt};
			\addlegendentry{$D'(\Sigma^*)$}
			\end{axis}
			\end{tikzpicture}
			\caption{Evolution of the partial KL divergence as the dimension increases, with the optimal covariance matrix $\Sigma^*$ (blue circles), the sample covariance $\hat{\Sigma}^*$ (red squares), and the projected covariance $\hat\Sigma^*_k$ (black triangles).}
			\label{fig:inefficiency}
		\end{subfigure}
		\hfill
		\begin{subfigure}[t]{.45\textwidth}
			\centering
			\begin{tikzpicture}[scale=.85]
			\begin{axis}[
			 xmin=0, ymin=0,
			legend style={at={(.5,-.1)}, anchor=north, draw=none}, grid, legend columns=2
			]
			\addplot [only marks, blue, mark=x, mark options={fill=blue}] table {Data2/Valp_sum.txt};
			\addlegendentry{$\hat \Sigma^*$}
			\addplot [only marks, red, mark=square, mark options={fill=red}] table {Data2/Valp_sum_th.txt};
			\addlegendentry{$\Sigma^*$}
			\end{axis}
			\end{tikzpicture}
			\caption{Computation of $\ell(\lambda_i)$ for the eigenvalues of $\Sigma^*$ (red squares) and $\hat \Sigma^*$ (blue crosses) in dimension $n = 100$.}
			\label{fig:eigsum}
		\end{subfigure}
		\caption{Partial KL divergence and spectrum for the function $\phi = \II_{\varphi \geq 0}$ with $\varphi$ the linear function given by~\eqref{eq:sum}.}
		\label{fig:eigsum+inefficiency}
	\end{figure}

		Figure \ref{fig:inefficiency}
	represents the evolution as the dimension varies between $5$ and $100$ of the partial KL divergence $D'$ for three different choices of covariance matrix: the optimal matrix $\Sigma^*$, its empirical estimation $\hat \Sigma^*$ and the estimation $\hat \Sigma^*_k$ of the optimal lower-dimensional covariance matrix. We can notice that the partial KL divergence for $\hat \Sigma^*$ grows much faster than the other two, and that the partial KL divergence for $\hat \Sigma^*_k$ remains very close to the optimal value $D'(\Sigma^*)$. As the KL divergence is a proxy for the efficiency of the auxiliary density (it is for instance closely related to the number of samples required for a given precision~\cite{Chatterjee18:0}), this suggests that using $\hat \Sigma^*_k$ will provide results close to optimal.

	We now check this claim. As $\Sigma^* = (v-1) \textbf{1} \textbf{1}^\top + I_n$, its eigenpairs are $(v, \textbf{1})$ and $(1,\dd_i)$ where the $\dd_i$'s form an orthonormal basis of the space orthogonal to the space spanned by~$\textbf{1}$. In particular, $(v, \textbf{1})$ is the largest (in $\ell$-order) eigenpair of~$\Sigma^*$ and $\Sigma^*_k = \Sigma^*$ for any $k \geq 1$. 
	%
	%
	%
	%
	In practice, we do not use this theoretical knowledge and $\Sigma^*$, $\Sigma^*_k$ and the eigenpairs are estimated. The six covariance matrices introduced in Section~\ref{def_cov} and in which we are interested are as follows:
	\begin{itemize}
		\item $\Sigma^* = (v-1) \textbf{1} \textbf{1}^\top + I_n$;
		\item $\hat \Sigma^*$ given by~\eqref{eq:hatSigma};
		\item $\Sigmatrois$ and $\Sigmacinq$ are equal and given by $(\hat \lambda-1) \textbf{1} \textbf{1}^\top + I_n$ with $\hat \lambda = \textbf{1}^\top \hat \Sigma^* \textbf{1}$. This amounts to assuming that the projection direction $\textbf{1}$ is perfectly known, whereas the variance in this direction is estimated;
		\item $\Sigmaquatre = (\hat \lambda - 1) \hat \dd {\hat \dd}^\top + I_n$ with $(\hat \lambda, \hat \dd)$ the smallest eigenpair of $\hat \Sigma^*$. The difference with the previous case is that we do not assume anymore that the optimal projection direction~$\textbf{1}$ is known, and so it needs to be estimated;
		\item $\Sigmasix = (\hat \lambda - 1) \frac{\hat \mm^* {(\hat \mm^*)}^\top}{\lVert \hat \mm^* \rVert^2} + I_n$ with $\hat \mm^*$ given by~\eqref{eq:hatm} and $\hat \lambda = \frac{{(\hat \mm^*)}^\top \hat \Sigma^* \hat \mm^*}{\lVert \hat \mm^* \rVert^2}$. Here we assume that $\mm^*$ is a good projection direction, but is unknown and therefore needs to be estimated.
	\end{itemize}
	Note that in the particularly simple case considered here, both $\hat \mm^* / \lVert \hat \mm^* \rVert$ and $\hat \dd$ are estimators of $\textbf{1}$ but they are obtained by different methods. In the next example we will consider a case where $\mm^*$ is not an optimal projection direction as given by Theorem~\ref{thm1}.
	
	Figure~\ref{fig:eigsum}
	%
	%
	represents the images by $\ell$ of the eigenvalues of $\Sigma^*$ and $\hat \Sigma^*$. This picture carries a very important insight. We notice that the estimation of most eigenvalues is poor: indeed, all the blue crosses except the leftmost one are meant to be estimator of~$1$, whereas we see that they are more or less uniformly spread between $0.4$ and $1.8$. This means that the variance terms in the corresponding directions are poorly estimated, which could be the explanation on why the use of $\hat \Sigma^*$ gives an inaccurate estimation. But what we remark also is that the function $\ell$ is quite flat around one: as a consequence, although the eigenvalues offer significant variability, this variability is smoothed by the action of $\ell$. Indeed, the images of the eigenvalues by $\ell$ take values between $0$ and $0.4$ and have smaller variability. Moreover, $\ell(x)$ increases sharply as $x$ approaches $0$ and thus efficiently distinguishes between the two leftmost estimated eigenvalues and is able to separate them.
	
	
	\subsubsection{Numerical results}

	We report in Table~\ref{table:sum} the numerical results for the six different matrices and the vMFN model for the dimension $n=100$. The column $\Sigma^*$ gives the optimal results, while the column $\hat \Sigma^*$ corresponds to the results that we are trying to improve. Comparing these two columns, we notice as expected that the estimation of $E$ with $\hat \Sigma^*$ is significantly degraded. Compared to the first column $\Sigma^*$, the third and fourth column with $\Sigmatrois =  \Sigmacinq$ correspond to the best projection direction $\textbf{1}$ (as for $\Sigma^*$) but estimating the variance in this direction (instead of the true variance) with $\textbf{1}^\top \hat \Sigma^* \textbf{1}$. This choice performs very well, with numerical results similar to the optimal ones. This can be understood since in this case, both $ \Sigmatrois$ and $\Sigma^*$ are of the form $\alpha \textbf{1} \textbf{1}^\top + I_n$ and so estimating $ \Sigmatrois$ requires only a one-dimensional estimation (namely, the estimation of $\alpha$). Next, the last two columns $\Sigmaquatre$ and $\Sigmasix$ highlight the impact of having to estimate the projection directions in addition to the variance since these two matrices are of the form $\hat \alpha \hat{\textbf{1}} {\hat{\textbf{1}}}^\top + I_n$ with both $\hat \alpha$ (the variance term) and $\hat{\textbf{1}}$ (the direction) being estimated. We observe that these matrices yield results which are close to optimal and greatly improve the estimation obtained using $\hat \Sigma^*$. In dimension $100$, the coefficient of variation is around~$4$~\% for $\Sigmasix$, and around~$5$~\% for $ \Sigmaquatre$, compared to $2.5$~\% for $\hat\Sigma^*$. 
\begin{table}[h]
		\begin{center}
			\small
			\begin{tabular}{V{3}lV{3}c|cV{3}c|cV{3}c|cV{3}cV{3}}
				\specialrule{.15em}{0em}{0em}
				 & $\Sigma^*$ & $\hat \Sigma^*$ & $\Sigmatrois$  & $\Sigmacinq$ & $\Sigmaquatre$ & $\Sigmasix$ & vMFN \\
				\specialrule{.15em}{0em}{0em}
				$D'(\Sigma) $ & 97 & 112 & 97 & 97 & 98 & 98  & / \\
				\hline
				Relative error ($\%$) & -0.6 & -26 & -0.3 & -0.3 & 0.0 & -0.4 & -0.5 \\
				\hline
				Coefficient of variation ($\%$) & 2.5 & 90 & 2.3 & 2.3 & 5.1 & 3.7 & 4.1 \\
				\specialrule{.15em}{0em}{0em}
			\end{tabular}
		\end{center}
		\caption{Numerical comparison of the estimation of $E \approx 1.35\cdot 10^{-3}$ considering the Gaussian model with the six covariance matrices defined in Section~\ref{def_cov} and the vFMN model, when $\phi = \II_{\{\varphi\geq 0\}}$ with $\varphi$ the linear function given by~\eqref{eq:sum}. As explained in the text, $\Sigmacinq$ and $\Sigmatrois$ are actually equal in this case.}
		\label{table:sum}
	\end{table}
	\color{black}
	Moreover, we observe that $\Sigmasix$ gives slightly better results than $\Sigmaquatre$. We suggest that this is because $\hat{\mm}^* / \lVert \hat{\mm}^* \rVert$ is a better estimator of $\textbf{1}$ than the eigenvector of $\hat{\Sigma}^*$. Indeed, evaluating $\hat{\mm}^*$ requires the estimation of $n$ parameters, whereas $\hat{\Sigma}^*$ needs around $n^2/2$ parameters to estimate, so the eigenvector is finally more noisy than the mean vector. 
	In the last column, we present the vMFN results that are similar to the estimation obtained with~$\Sigmasix$. \color{black} 
	
	Thus, the proposed idea improves significantly the probability estimation in high dimension. But we see that the method taken in \cite{MasriEtAl_ImprovementCrossentropyMethod_2020} with the projection~$\mm^*$ is at least as much efficient in this example where we need only a one-dimensional projection. The next case shows that the projection on more than one direction can outperform the one-dimensional projection on~$\mm^*$. 

	\subsection{Test case 2: projection in 2 directions}\label{sub:parabol}
	
	The second test case is again a probability estimation, i.e., it is of the form $\phi = \II_{\{\varphi \geq 0\}}$ with now the function $\varphi$ having some quadratic terms:
	\begin{align}\label{eq:parabol}
	\varphi: \xx=(x_1,\ldots,x_n) \in \RR^n \mapsto x_1 - 25 x_2^2 - 30 x_3^2 - 1.
	\end{align}
		The quantity of interest $E$ is defined as $E=\int_{\mathbb{R}^n} \phi(\xx) f(\xx) \textrm{d}\xx = \PP_f(\varphi(\XX)\geq 0)$ for all $n$ where the density $f$ is the standard $n$-dimensional Gaussian distribution. \color{black}This function is motivated in part because $\mm^*$ and $\dd^*_1$ are different and also because Algorithm~\ref{algo:choicek} chooses two projection directions. Thus, this is an example where $\Sigmacinq$ and $\Sigmatrois$ are significantly different.

	\subsubsection{Evolution of the partial KL divergence and spectrum}

\begin{figure}[h]
		\begin{subfigure}[t]{.45\textwidth}
			\centering
			\begin{tikzpicture}[scale=.9]
			\begin{axis}[
			xlabel={dimension $n$},
			ylabel={},
			xmin=0,ymin=-10,
			legend style={at={(.5,-.1)}, anchor=north, draw=none, font=\small}, grid, legend columns = 3
			]
			\addplot [ red, mark=square*, mark options={fill=red}] table {Data2/Dkl1_parabz.txt};
			\addlegendentry{$D'(\hat \Sigma^*)$}
			
			\addplot [ black, mark=triangle*, mark options={fill=black}] table {Data2/Dklproj_parabz.txt};
			\addlegendentry{$D'(\hat \Sigma^*_k)$}
			\addplot [ blue, mark=*, mark options={fill=blue}] table {Data2/Dklstar_parabz.txt};
			\addlegendentry{$D'(\Sigma^*)$}
			\end{axis}
			\end{tikzpicture}
			\caption{Evolution of the partial KL divergence as the dimension increases, with the optimal covariance matrix $\Sigma^*$ (blue circles), the sample covariance $\hat{\Sigma}^*$ (red squares), and the projected covariance $\hat\Sigma^*_k$ (black triangles).}
			\label{fig:inefficiency-parab}
		\end{subfigure}
		\hfill
		\begin{subfigure}[t]{.45\textwidth}
			\centering
			\begin{tikzpicture}[scale=.9]
			\begin{axis}[
			xmin=0,ymin=0,
			legend style={at={(.5,-.1)}, anchor=north, draw=none}, grid, legend columns=2
			]
			\addplot [only marks, blue, mark=x, mark options={fill=blue}] table {Data2/Valp_parabz.txt};
			\addlegendentry{$\hat \Sigma^*$}
			\addplot [only marks, red, mark=square, mark options={fill=red}] table {Data2/Valp_parab_th.txt};
			\addlegendentry{$\Sigma^*$}
			\end{axis}
			\end{tikzpicture}
			\caption{Computation of $\ell(\lambda_i)$ for the eigenvalues of $\Sigma^*$ (red squares) and $\hat \Sigma^*$ (blue crosses) in dimension $n = 100$.}
			\label{fig:eigsum-parab}
		\end{subfigure}
		\caption{Partial KL divergence and spectrum $\phi = \II_{\varphi \geq 0}$ with $\varphi$ given by~\eqref{eq:parabol} in dimension $n=100$. Left: same behavior as for the first test case. Right: we now have two eigenvalues that stand out, and the behavior of $\ell$ is such that Algorithm~\ref{algo:choicek} selects $k = 2$ which corresponds to the leftmost two. 
		}
		\label{fig:eigparabol}
	\end{figure}
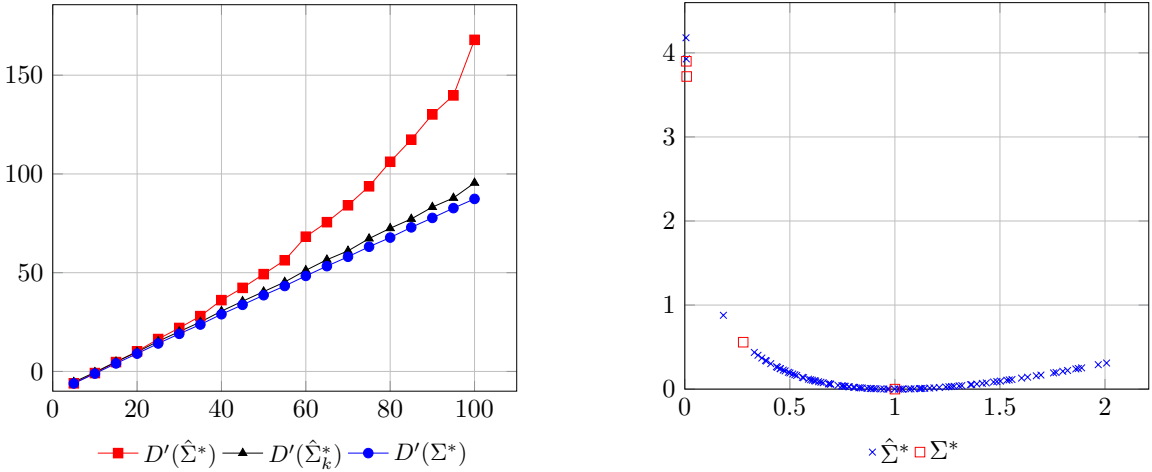

	We check on Figure~\ref{fig:inefficiency-parab}
	that the partial KL divergence obeys the same behavior as for the previous example, namely the one associated with $\hat \Sigma^*$ increases much faster than the ones associated with $\Sigma^*$ and $\hat \Sigma^*_k$, which again suggests that projecting can improve the situation. Since the function $\varphi$ only depends on the first three variables and is even in $x_2$ and~$x_3$, one gets that $\mm^* = \alpha 
	\textbf{e}_1$ with $\alpha = \EE(X_1 \mid X_1 \geq 25 X^2_2 + 30 X^2_3 + 1) \approx 1.9$ (here and in the sequel, $\textbf{e}_i$ denotes the $i$th canonical vector of $\RR^n$, i.e., all its coordinates are $0$ except the $i$-th one which is equal to one), and that~$\Sigma^*$ is diagonal with
	\[ \Sigma^* =
	\begin{pmatrix}
	\lambda_1 & 0 & 0 & 0 & \cdots & 0 \\
	0 & \lambda_2 & 0 & 0 & \cdots & 0 \\
	0 & 0 & \lambda_3 & 0 & \cdots & 0 \\
	0 & 0 & 0 & 1 & \cdots & 0 \\
	\vdots & \vdots & \vdots & \vdots & \ddots & \vdots \\
	0 & 0 & 0 & 0 & \cdots & 1 \\
	\end{pmatrix}. \]
	Note that the off-diagonal elements of the submatrix $(\Sigma^*_{ij})_{1 \leq i, j \leq 3}$ are indeed $0$ since they result from integrating an odd function of an odd random variable with an even conditioning. For instance, if $F(x) = \PP(30 X^2_3 + 1 \leq x)$, then by conditioning on $(X_1, X_3)$ we obtain
	\begin{align*}
	\Sigma^*_{12} & = \EE \left( (X_1 - \alpha) X_2 \mid X_1 - 25 X_2^2 \geq 30 X^2_3 + 1 \right)\\
	& = \frac{1}{E} \EE \left[ (X_1 - \alpha) \EE \left( X_2 F(X_1 - 25 X^2_2) \mid X_1 \right) \right]
	\end{align*}
	which is $0$ as $x_2 F(x_1 - x^2_2)$ is an odd function of $x_2$ for fixed $x_1$, and $X_2$ has an even density. 
	We can numerically compute $\lambda_1 \approx 0.278$, $\lambda_2 \approx 0.009$ and $\lambda_3 \approx 0.0075$. These values correspond to the red squares in Figure~\ref{fig:eigsum-parab} which shows that the smallest eigenvalues are properly estimated. Moreover, Algorithm~\ref{algo:choicek} selects the two largest eigenvalues, which have the highest $\ell$-values. These two eigenvalues thus correspond to the eigenvectors $\mathbf{e}_2$ and $\mathbf{e}_3$, and so we see that on this example, the optimal directions predicted by Theorem~\ref{thm1} are significantly different (actually, orthogonal) from~$\mm^*$ which is proportional to $\textbf{e}_1$.

	\subsubsection{Numerical results}

	The numerical results of our simulations are presented in Table~\ref{table:parabol}. We remark as before that, when using $\hat \Sigma^*$, the accuracy quickly deteriorates as the dimension increases as shows the coefficient of variation of $84~\%$ in dimension $n = 100$. In contrast, $\Sigmatrois$ leads to very accurate results, which remain close to optimal up to the same dimension $n = 100$. This behavior is to compare with the evolution of the relative KL divergence: contrary to $\hat \Sigma^*$,~$\Sigmatrois$ gives a partial KL divergence close to optimal in dimension $n = 100$. This confirms that the KL divergence is indeed a good proxy to assess the relevance of an auxiliary density.
	\begin{table}[h]
		\begin{center}
			\small
			\begin{tabular}{V{3}lV{3}c|cV{3}c|cV{3}c|cV{3}cV{3}}
				\specialrule{.15em}{0em}{0em}
			 & $\Sigma^*$ & $\hat{\Sigma}^*$ & $\Sigmatrois$ & $\Sigmacinq$ & $\Sigmaquatre$ & $\Sigmasix$ & vMFN \\
				\specialrule{.15em}{0em}{0em}
				$D'(\Sigma) $ & 89 & 104 & 90 & 97 & 90 & 97 & / \\
				\hline
				Relative error ($\%$) & -0.8 & -27 & -0.4 & 0.2 & -1.7 & -1.7 & 2.8 \\
				\hline
				Coefficient of variation ($\%$) & 2.6 & 84 & 2.9 & 25 & 7.0 & 22 & 30 \\
				\specialrule{.15em}{0em}{0em}
			\end{tabular}
		\end{center}
		\caption{Numerical comparison of the estimation of $E \approx 1.51\cdot 10^{-3}$ considering the Gaussian density with the six covariance matrices defined in Section \ref{def_cov} and the vFMN density when $\phi = \II_{\{\varphi\geq 0\}}$ with $\varphi$ the quadratic function given by~\eqref{eq:parabol}.}
		\label{table:parabol}
	\end{table}
	It is also interesting to note that the direction~$\mm^*$ improves the situation compared to not projecting (column $\Sigmacinq$ compared to~$\hat \Sigma^*$), but using~$\Sigmatrois$ gives significantly better results, with for instance a coefficient of variation around~$3~\%$ for $\Sigmatrois$ and around $25~\%$ for $\Sigmacinq$ in dimension $n = 100$. Thus, this confirms our theoretical result that the $\dd^*_i$'s are good directions on which to project. 
	
	Finally, we notice that performing estimations of the projection directions instead of taking the true ones (columns $\Sigmaquatre$ vs~$\Sigmatrois$) slightly degrades the situation, making the coefficient of variation increase from $3$ to~$7~\%$ even if the accuracy remains satisfactory. 
	The vMFN model is also not really adapted to this example as it gives results similar to $ \Sigmacinq$. Gaussian density family are more able to fit $g^*$ than vMFN parametric model in this test case.  \color{black}

	\begin{rem} \label{rem:FIS}
		For the two text cases studied so far, projecting $\hat \Sigma^*$ in the Failure-Informed Subspace (FIS) of~\cite{UribeEtAl_CrossentropybasedImportanceSampling_2020} (see the introduction) would outperform our method with $\hat \Sigma^*_k$, leading to results close to those obtained with $\Sigma^*$. However, computing the FIS relies on the knowledge of the gradient of the function $\varphi$, which is straightforward to compute in these two test cases, and the method of~\cite{UribeEtAl_CrossentropybasedImportanceSampling_2020} can be applied because they are rare-event problems (i.e., $\phi$ is of the form $\phi = \II_{\{\varphi \geq 0\}}$). In the next three sections, we present two applications in mathematical finance where the evaluation of the FIS is not feasible since either the function is not differentiable (test case of Section~\ref{sub:portfolio}) or the example is not a rare event simulation problem (test cases of Sections~\ref{sub:banana} and \ref{sub:payoff}).
	\end{rem}

		\subsection{Test case 3: Banana shape distribution}\label{sub:banana} 
\begin{figure}[h]
	
				\color{black}
		\begin{subfigure}[t]{.45\textwidth}
			\centering
			\begin{tikzpicture}[scale=.9]
			\begin{axis}[
			xlabel={dimension $n$},
			ylabel={},
			xmin=0,ymin=-10,
			legend style={at={(.5,-.1)}, anchor=north, draw=none, font=\small}, grid, legend columns = 3
			]
			\addplot [ red, mark=square*, mark options={fill=red}] table {Data2/Dkl1_bananab.txt};
			\addlegendentry{$D'(\hat \Sigma^*)$}
			
			\addplot [ black, mark=triangle*, mark options={fill=black}] table {Data2/Dklproj_bananab.txt};
			\addlegendentry{$D'(\hat \Sigma^*_k)$}
			\addplot [ blue, mark=*, mark options={fill=blue}] table {Data2/Dklstar_bananab.txt};
			\addlegendentry{$D'(\Sigma^*)$}
			\end{axis}
			\end{tikzpicture}
			\caption{
			Evolution of the partial KL divergence as the dimension increases, with the optimal covariance matrix $\Sigma^*$ (blue circles), the sample covariance $\hat{\Sigma}^*$ (red squares), and the projected covariance $\hat\Sigma^*_k$ (black triangles).}
			\label{fig:inefficiency-banana}
		\end{subfigure}
		\hfill
		\begin{subfigure}[t]{.45\textwidth}
			\centering
			\begin{tikzpicture}[scale=.9]
			\begin{axis}[
			legend style={at={(.5,-.1)}, anchor=north, draw=none}, grid, legend columns=2, xmin=0, ymin=0,
			]
			\addplot [only marks, blue, mark=x, mark options={fill=blue}] table {Data2/Valp_banana_d100.txt};
			\addlegendentry{$\hat \Sigma^*$}
			\addplot [only marks, red, mark=square, mark options={fill=red}] table {Data2/Valpstar_banana_d100.txt};
			\addlegendentry{$\Sigma^*$}
			\end{axis}
			\end{tikzpicture}
		\caption{
		Computation of $\ell(\lambda_i)$ for the eigenvalues of $\Sigma^*$ (red squares) and $\hat \Sigma^*$ (blue crosses) in dimension $n = 100$ for the banana shape example of~\eqref{eq:banana}.
		}
			\label{fig:eigbanana}
		\end{subfigure}
		\caption{
		Partial KL divergence and spectrum for the banana shape example.}
		\label{fig:eigbanana+inefficiency}
	\end{figure}
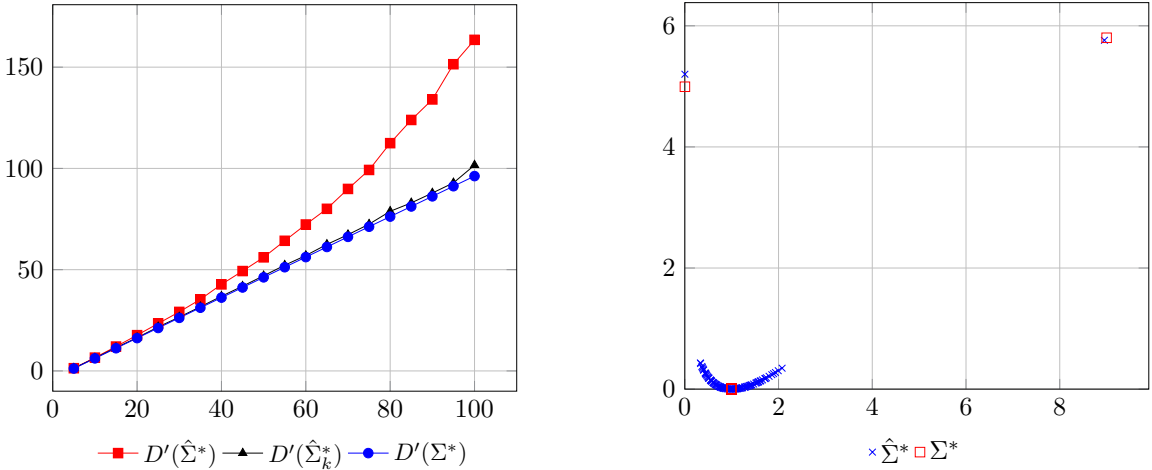

		The third test case we consider is the integration of the banana shape distribution $h$, which is a classical test case in importance sampling \cite{cornuet2012adaptive, ElviraEtAl_GeneralizedMultipleImportance_2019}. The banana shape distribution is the following pdf
	\begin{align} \label{eq:banana}
	h(\xx) = g_{{\bf 0},C}(x_1,x_2+b(x_1^2-\sigma^2),x_3,\dots,x_n).
	\end{align}
 The term $g_{{\bf 0},C}$ represents the pdf of a Gaussian distribution of mean ${\bf 0}$ and diagaonal covariance matrix $C=\text{diag}(\sigma^2,1,\dots,1)$. The value of $b$ and $\sigma^2$ are respectively set to $b=800$ and $\sigma^2=0.0025$. We choose $\phi$ such that the optimal IS density~$g^*$ is equal to $h$, i.e., we choose $\phi = h/f$ so that the integral $E$ that we are trying to estimate is equal to $E = \int \phi f = 1$. This choice is made in order to have an optimal covariance matrix $\Sigma^*$ whose two largest eigenvalues (in $\ell$-order) correspond to the smallest and largest eigenvalues, as can be seen in Figure~\ref{fig:eigbanana}. More formally, the optimal value of the Gaussian parameters are given by $\mm^*={\bf 0}$ and $\Sigma^*$ is diagonal with
	\[ \Sigma^* =
	\begin{pmatrix}
	0.0025 & 0 & 0 & 0 & \cdots & 0 \\
	0 & 9 & 0 & 0 & \cdots & 0 \\
	0 & 0 & 1 & 0 & \cdots & 0 \\
	0 & 0 & 0 & 1 & \cdots & 0 \\
	\vdots & \vdots & \vdots & \vdots & \ddots & \vdots \\
	0 & 0 & 0 & 0 & \cdots & 1 \\
	\end{pmatrix}. \]
	%
	The evolution of the KL partial divergence is given in Figure~\ref{fig:inefficiency-banana}.
	As the optimal mean $\mm^*$ is equal to ${\bf 0}$, we cannot project on $\mm^*$ and so the matrix $\Sigmacinq$ is not defined. However, the numerical estimation $\hat \mm^*$ will not be equal to $0$ and so the approach proposed in \cite{MasriEtAl_ImprovementCrossentropyMethod_2020} with $\Sigmasix$ is still applicable numerically.\\
	The simulation results for the different covariance matrices and the vMFN density are given in Table \ref{table:banana}. The matrices $\Sigmatrois$ and $\Sigmaquatre$ perform very well for the estimation of $E$ with an accuracy of the same order as the optimal covariance matrix $\Sigma^*$. The effect of estimating the $k=2$ main projection directions does not affect much the estimation performance as $\Sigmaquatre$ is still efficient compared to $\Sigmatrois$. The estimation results with $\Sigmasix$ are not really accurate and this choice is in fact roughly equivalent to choosing a random projection direction. The vMFN parametric model is not adapted to this test case as the vMFN estimate is not close to 1.
	\color{black}

		\begin{table}[h]
		\begin{center}
			\small
			\begin{tabular}{V{3}lV{3}c|cV{3}c|cV{3}c|cV{3}cV{3}}
				\specialrule{.15em}{0em}{0em}
			 & $\Sigma^*$ & $\hat{\Sigma}^*$ & $\Sigmatrois$ & $\Sigmacinq$ & $\Sigmaquatre$ & $\Sigmasix$ & vMFN \\
				\specialrule{.15em}{0em}{0em}
				$D'(\Sigma) $ &  96 & 111 & 96 & NA & 97 & 107 & / \\
				\hline
				Relative error ($\%$) & -0.2 & -38 & -2.4 & NA & -3.1 & -11 & -17 \\
				\hline
				 Coefficient of variation ($\%$) & 6.1 & 74 & 5.8 & NA & 7.6 & 25 & 32 \\
				\specialrule{.15em}{0em}{0em}
			\end{tabular}
		\end{center}
				\caption{
				Numerical comparison of the estimation of $E=1$ considering the banana shape density with the four covariance matrices defined in Section \ref{def_cov} and the vMFN density when $\phi = h/f$. NA stands for non applicable, as explained in the text.}
			\label{table:banana}
	\end{table}
	
	\color{black}
	
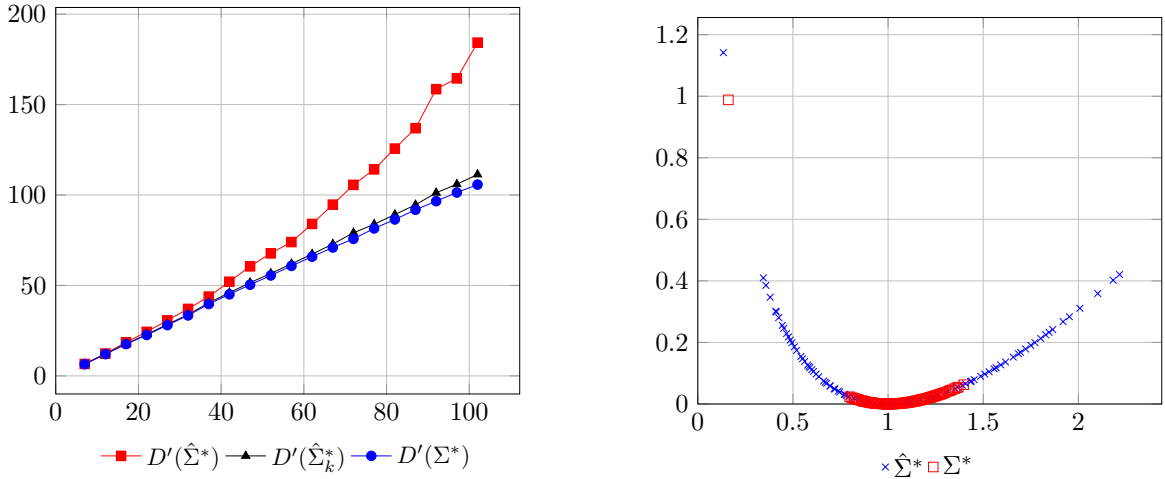
\begin{figure}[h]
			\begin{subfigure}{.45\textwidth}
			\centering
			\begin{tikzpicture}[scale=.9]
			\begin{axis}[
			xlabel={dimension $n$},
			ylabel={},
			xmin=0,ymin=-10,
			legend style={at={(.5,-.1)}, anchor=north, draw=none, font=\small}, grid, legend columns = 3
			]
			\addplot [ red, mark=square*, mark options={fill=red}] table {Data2/Dkl1_portfolio.txt};
			\addlegendentry{$D'(\hat \Sigma^*)$}
			
			\addplot [ black, mark=triangle*, mark options={fill=black}] table {Data2/Dklproj_portfolio.txt};
			\addlegendentry{$D'(\hat \Sigma^*_k)$}
			\addplot [ blue, mark=*, mark options={fill=blue}] table {Data2/Dklstar_portfolio.txt};
			\addlegendentry{$D'(\Sigma^*)$}
			\end{axis}
			\end{tikzpicture}
			\caption{Evolution of the partial KL divergence as the dimension increases, with the optimal covariance matrix $\Sigma^*$ (blue circles), the sample covariance $\hat{\Sigma}^*$ (red squares), and the projected covariance $\hat\Sigma^*_k$ (black triangles).}
			\label{fig:inefficiency-portfolio}
		\end{subfigure}
		\hfill
		\begin{subfigure}{0.45\textwidth}
		\centering
			\begin{tikzpicture}[scale=.9]
			\begin{axis}[
			legend style={at={(.5,-.1)}, anchor=north, draw=none}, grid, legend columns=2, xmin=0, ymin=0
			]
			\addplot [only marks, blue, mark=x, mark options={fill=blue}] table {Data2/Valp_portfolio_d100.txt};
			\addlegendentry{$\hat \Sigma^*$}
			\addplot [only marks, red, mark=square, mark options={fill=red}] table {Data2/Valpstar_portfolio_d100.txt};
			\addlegendentry{$\Sigma^*$}
			\end{axis}
			\end{tikzpicture}
		\caption{ 
		Computation of $\ell(\lambda_i)$ for the eigenvalues of $\Sigma^*$ (red squares) and $\hat \Sigma^*$ (blue crosses) in dimension $n = 100$ for the large portfolio losses of \eqref{eq:portfolio}. 
		}
			\label{fig:eigportfolio}
		\end{subfigure}
		\caption{ 
		Partial KL divergence and spectrum for the function $\phi = \II_{\varphi \geq 0}$ with $\varphi$ the function given by~\eqref{eq:portfolio}.}
		\label{fig:eigportfolio+inefficiency}
	\end{figure}

The next example is a rare event application in finance, taken from \cite{BassambooEtAl_PortfolioCreditRisk_2008, ChanKroese_ImprovedCrossentropyMethod_2012}. 
	The unknown integral is $E=\int_{\mathbb{R}^{n+2}} \phi(\xx) f(\xx) \textrm{d}\xx = \PP_f(\varphi(\XX)\geq 0)$, with $\phi = \II_{\{\varphi \geq 0\}}$ and $f$ is the standard $n+2$-dimensional Gaussian distribution. The function $\varphi$ is the portfolio loss function defined as:
	\begin{align} \label{eq:portfolio}
	\varphi(\xx) = \underset{j=3}{\overset{n+2}{\sum}} \II_{\{\Psi(x_1, x_2, x_j) \geq 0.5\sqrt{n}\}}-0.25 n,
	\end{align}
	with
	\[ \Psi(x_1, x_2, x_j) = \left( q x_1 + 3 (1-q^2)^{1/2}x_j \right) \left[ F_\Gamma^{-1} \left( F_\NO({x_2}) \right) \right]^{-1/2}, \]
	 where $F_\Gamma$ and $F_\NO$ are the cumulative distribution functions of $\text{Gamma}(6,6)$ and $\NO(0,1)$ random variables respectively.\color{black}\\
	The reference value of this probability $E$ is reported in Table \ref{table:portfolio} for dimension $n=100$. The optimal parameters $\mm^*$ and $\Sigma^*$ cannot be computed analytically, but they are accurately estimated by Monte Carlo with a large sample. It turns out that $\mm^*$ and the first eigenvector $\dd^*_1$ of $\Sigma^*$ are numerically indistinguishable 
	and that Algorithm~\ref{algo:choicek} selects $k=1$ projection direction, so that numerically, the choices $\Sigmatrois$ and $\Sigmacinq$ are indistinguishable and gives exactly the same estimation results. The KL partial divergence and the spectrum with the associated $\ell$-order are presented respectively in Figure \ref{fig:inefficiency-portfolio} and in Figure  \ref{fig:eigportfolio}.

		\begin{table}[h]
		\begin{center}
			\small
				\begin{tabular}{V{3}lV{3}c|cV{3}c|cV{3}c|cV{3}cV{3}}
				\specialrule{.15em}{0em}{0em}
			 & $\Sigma^*$ & $\hat \Sigma^*$ & $\Sigmatrois$ & $\Sigmacinq$ & $\Sigmaquatre$ & $\Sigmasix$  & vMFN \\
				\specialrule{.15em}{0em}{0em}
				$D'(\Sigma) $ & 106 & 122 & 107 & 107 & 108 & 107 & /\\
				\hline
				Relative error ($\%$) & -1.9 & -37 & 2.5 & 2.5 & 1.0 & -0.3 & 0.7 \\
				\hline
                Coefficient of variation ($\%$) & 11 & 73 & 11 & 11 & 17 & 13 & 6.1 \\
				\specialrule{.15em}{0em}{0em}
			\end{tabular}
		\end{center}
		\caption{Numerical comparison of the estimation of $E \approx 1.82 \cdot 10^{-3}$ using the six covariance matrices of Section~\ref{def_cov} and the vFMN density when $\phi = \II_{\{\varphi \geq 0\}}$ with $\varphi$ given by~\eqref{eq:portfolio}.}
		\label{table:portfolio}
	\end{table}

	The results of Table~\ref{table:portfolio} show similar trends as for the first test case of Section~\ref{sub:sum}. First, projecting seems indeed a relevant idea, as using $\Sigmatrois$ or $\Sigmacinq$ greatly improves the situation compared to $\hat \Sigma^*$. This is particularly salient as $\hat \Sigma^*$ yields an important bias and a coefficient of variation of $94~\%$, whereas projecting on $\dd^*_1$ or $\mm^*$ yields a coefficient of variation of $11~\%$. This improvement is still true even when the projection directions are estimated: compared to $\Sigmatrois$ and $\Sigmacinq$, $\Sigmaquatre$ and $\Sigmasix$ give coefficients of variation between $16$ and $13~\%$. Finally, $\Sigmasix$ seems to behave better than $\Sigmaquatre$. As for the first test case, this can be understood by the the fact that $\hat \mm^*$ is probably a more accurate estimator of $\mm^*$ than $\hat \dd^*_1$ of $\dd^*_1$, as $\mm^*$ and $\dd^*_1$ are numerically indistinguishable.

	\subsection{Application: discretized Asian payoff }\label{sub:payoff}
	
\begin{figure}[h]
			\begin{subfigure}{.45\textwidth}
			\centering
			\begin{tikzpicture}[scale=.9]
			\begin{axis}[
			xlabel={dimension $n$},
			ylabel={},
			xmin=0,ymin=-10,
			legend style={at={(.5,-.1)}, anchor=north, draw=none, font=\small}, grid, legend columns = 3
			]
			\addplot [ red, mark=square*, mark options={fill=red}] table {Data2/Dkl1_payoff.txt};
			\addlegendentry{$D'(\hat \Sigma^*)$}
			
			\addplot [ black, mark=triangle*, mark options={fill=black}] table {Data2/Dklproj_payoff.txt};
			\addlegendentry{$D'(\hat \Sigma^*_k)$}
			\addplot [ blue, mark=*, mark options={fill=blue}] table {Data2/Dklstar_payoff.txt};
			\addlegendentry{$D'(\Sigma^*)$}
			\end{axis}
			\end{tikzpicture}
			\caption{ 
			Evolution of the partial KL divergence as the dimension increases, with the optimal covariance matrix $\Sigma^*$ (blue circles), the sample covariance $\hat{\Sigma}^*$ (red squares), and the projected covariance $\hat\Sigma^*_k$ (black triangles).}
			\label{fig:inefficiency-kawai}
		\end{subfigure}
		\hfill
		\begin{subfigure}{0.45\textwidth}
			\centering
			\begin{tikzpicture}[scale=.9]
			\begin{axis}[
			legend style={at={(.5,-.1)}, anchor=north, draw=none}, grid, legend columns=2, xmin=0, ymin=0
			]
			\addplot [only marks, blue, mark=x, mark options={fill=blue}] table {Data2/Valp_kawai_d100.txt};
			\addlegendentry{$\hat \Sigma^*$}
			\addplot [only marks, red, mark=square, mark options={fill=red}] table {Data2/Valpstar_kawai_d100.txt};
			\addlegendentry{$\Sigma^*$}
			\end{axis}
			\end{tikzpicture}
		\caption{ 
		Computation of $\ell(\lambda_i)$ for the eigenvalues of $\Sigma^*$ (red squares) and $\hat \Sigma^*$ (blue crosses) in dimension $n = 100$ for the Asian payoff example of \eqref{eq:payoff}
		}
			\label{fig:eigkawai}
		\end{subfigure}
		\caption{Partial KL divergence and spectrum for the function $\phi$ given in~\eqref{eq:payoff}}
		\label{fig:eigkawai+inefficiency}
	\end{figure}
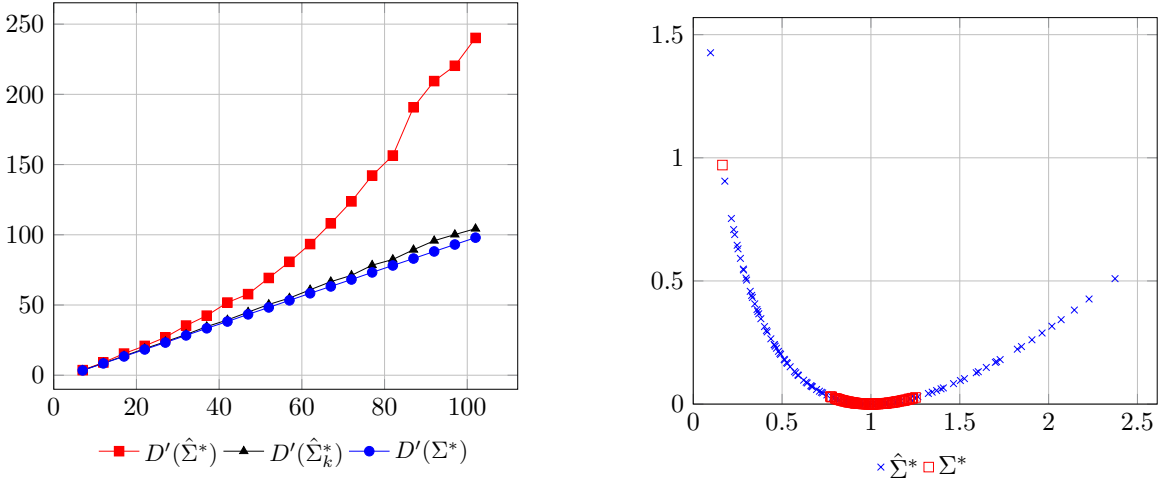

	Our last numerical experiment is a mathematical finance example coming from~\cite{Kawai_OptimizingAdaptiveImportance_2018}, representing a discrete approximation of a standard Asian payoff under the Black--Scholes model. 
	The goal is to estimate the integral $E=\int_{\mathbb{R}^n} \phi(\xx) f(\xx) \textrm{d}\xx$  with $f$ the standard $n$-dimensional Gaussian distribution \color{black}and the following function $\phi$: 
	\begin{align}\label{eq:payoff}
	\phi: \xx=(x_1,\ldots,x_n) \mapsto e^{-rT}\left[\frac{S_0}{n} \sum_{i=1}^n \exp\left(\sum_{k=1}^{i} \left(r-\frac{\sigma^2}{2}\right)\frac{T}{n}+\sigma \sqrt{\frac{T}{n}}x_k \right)-K\right]_+
	\end{align}
	where $[y]_+=\max(y,0)$, for a real number $y$. The constants are taken from~\cite{Kawai_OptimizingAdaptiveImportance_2018}: $S_0=50$, $r=0.05, T=0.5, \sigma=0.1, K=55$, where they test the function for dimension $n=16$. In our contribution, we test this example in dimension $100$. Concerning $\mm^*$ and the $\dd^*_i$'s, the situation is the same as in the previous example: they are not available analytically but can be estimated numerically by Monte Carlo with a large simulation budget. And again, it turns out that $\mm^*$ and the first eigenvector $\dd^*_1$ of $\Sigma^*$ are numerically indistinguishable and that Algorithm~\ref{algo:choicek} selects $k=1$ projection direction, so that we have $\Sigmatrois$ and $\Sigmacinq$ yield results that are numerically indistinguishable. The KL partial divergence and the spectrum with the associated $\ell$-order are respectively presented in Figure \ref{fig:inefficiency-kawai} and Figure \ref{fig:eigkawai}. Although for the first test case of Section~\ref{sub:sum} it is clear why we should have $\mm^* = \dd^*_1$, we find it striking that in the two numerical examples coming from real-world applications we also have this. We will elaborate on this in the conclusion.
	
	The results of this example are given in Table \ref{table:payoff}. 
	\begin{table}[h]
		\begin{center}
			\small
\begin{tabular}{V{3}lV{3}c|cV{3}c|cV{3}c|cV{3}cV{3}}
				\specialrule{.15em}{0em}{0em}
			 & $\Sigma^*$ & $\hat \Sigma^*$ & $ \Sigmatrois$ & $\Sigmacinq$ & $\Sigmaquatre$ & $\Sigmasix$ & vMFN \\
			 				\specialrule{.15em}{0em}{0em}
				$D'(\Sigma) $ & 97 & 138 & 98 & 98 & 100 & 98 &  / \\
				\hline
				Relative error ($\%$) & 0.2 & -94 & 0.2 & 0.2 & -4.4 & -0.7 & 0.3 \\
				\hline
				Coefficient of variation ($\%$) & 3.1 & 94 & 2.6 & 2.6 & 9.1 & 2.5 & 2.4 \\
				\specialrule{.15em}{0em}{0em}
			\end{tabular}
		\end{center}
		\caption{Numerical comparison of the estimation of $E \approx 18.7 \times 10^{-3}$ using the six covariance matrices of Section~\ref{sub:main-result+positioning} and the vFMN density when $\phi$ is given by~\eqref{eq:payoff}.}
		\label{table:payoff}
	\end{table}
	The insight gained in the previous examples is confirmed. Projecting on $\mm^*$ or $\dd^*_1$ in dimension $n = 100$ enables to reach convergence and reduces (compared to $\hat \Sigma^*$) the coefficient of variation from $94~\%$ to $2.6~\%$. Moreover, this improvement goes through even when projection directions are estimated, with again $\Sigmasix$ behaving better than $\Sigmaquatre$. 


	\section{Conclusion} \label{sec:Ccl}
	
	The goal of this paper is to assess the efficiency of projection methods in order to overcome the curse of dimensionality for importance sampling. Based on a new theoretical result (Theorem~\ref{thm1}), we propose to project on the subspace spanned by the eigenvectors $\dd^*_i$'s corresponding to the largest eigenvalues of the optimal covariance matrix~$\Sigma^*$, where eigenvalues are ranked based on their image by some explicit function $\ell$. Our numerical results show that if the $\dd^*_i$'s were perfectly known, then projecting on them (column~$\Sigmatrois$ in the result tables) would greatly improve the final estimation compared to using the empirical estimation of the covariance matrix (column~$\hat \Sigma^*$) and actually lead to results which are comparable to those obtained with the optimal covariance matrix (column $\Sigma^*$). Moreover, we show that this improvement goes through when one estimates the $\dd^*_i$'s (column $\Sigmaquatre$) by computing the eigenpairs of $\hat \Sigma^*$: indeed, using $\Sigmaquatre$ as covariance matrix instead of $\hat \Sigma^*$ gives results which remain accurate up to the dimension $n = 100$ considered in the present paper.
	
	Moreover, we compare the directions $\dd^*_i$'s, which are justified by Theorem~\ref{thm1}, with the algorithm proposed in~\cite{MasriEtAl_ImprovementCrossentropyMethod_2020} which amounts to projecting on $\mm^*$ (column~$\Sigmacinq$ when $\mm^*$ is assumed to be known, or column $\Sigmasix$ when one uses the estimation $\hat \mm^*$ of $\mm^*$). On three out of the five numerical examples considered, it turns out that $\mm^*$ and $\dd^*_1$ can be proved to be equal, or are numerically indistinguishable, and Algorithm~\ref{algo:choicek} selects one projection direction. In these cases, the choices $\Sigmatrois$ and $\Sigmacinq$ thereforce lead to similar numerical results. Although for the first example of Section~\ref{sub:sum} the relation $\mm^* = \dd^*_1$ is easy to get thanks to the high symmetry of the problem at hand, for the concrete applications of Sections~\ref{sub:portfolio} and~\ref{sub:payoff} this is more surprising. The second and third test cases of Section~\ref{sub:parabol} break the relation between $\mm^*$ and $\dd^*_1$, and in this case, $\Sigmaquatre$ clearly outperforms $\Sigmasix$. It would be interesting to delve deeper into the relation between $\mm^*$ and the eigenvectors of $\Sigma^*$, and try and understand when estimating the $\dd^*_i$'s instead of the simpler $\mm^*$ is indeed worthwhile.
	
	All in all, these theoretical and numerical results show that the $\dd^*_i$'s of Theorem~\ref{thm1} are good directions in which to estimate variance terms. With the insight gained, we see several ways to extend our results. Two in particular stand out:
	\begin{enumerate}
		\item study different ways of estimating the eigenpairs $(\lambda^*_i, \dd^*_i)$;
		\item incorporate this method in adaptive importance sampling schemes, in particular the cross-entropy method \cite{RubinsteinKroese_SimulationMonteCarlo_2017}.
	\end{enumerate}
	For the first point, remember that we made the choice to estimate the eigenpairs of $\Sigma^*$ by computing the eigenpairs of $\hat \Sigma^*$. Moreover, in the numerical examples of Sections~\ref{sub:sum},~\ref{sub:portfolio} and~\ref{sub:payoff} where $\mm^*$ and $\dd^*_1$ are equal or indistinguishable, we saw that $\Sigmasix$ performed better than $\Sigmaquatre$ and we conjecture that this is because $\hat \mm$ is a better estimator than $\hat \dd^*_1$. This suggests that improving the estimation of the $\dd^*_i$'s can indeed improve the final estimation of $E$. Possible ways to do so consist in adapting existing results on the estimation of covariance matrices (for instance~\cite{LedoitWolf_WellconditionedEstimatorLargedimensional_2004}) or even directly results on the estimation of eigenvalues of covariance matrices such as~\cite{Benaych-Georges11:0, Mestre_ImprovedEstimationEigenvalues_2008, Mestre_AsymptoticBehaviorSample_2008, Nadakuditi08:0}, which we plan to do in future work. Moreover, it would be interesting to relax the assumption that one can sample from $g^*$ in order to estimate~$\hat \Sigma^*$. For the second point, we plan to investigate how the idea of the present paper can improve the efficiency of adaptive importance sampling schemes in high dimension. In this case, there is an additional difficulty, namely the introduction of likelihood ratios can lead to the problem of weight degeneracy which is another reason why performance of such schemes degrades in high dimension~\cite{BengtssonEtAl_CurseofdimensionalityRevisitedCollapse_2008}.
	
	We note finally that it would be interesting to consider multimodal failure functions~$\phi$. Indeed, with unimodal functions, the light tail of the Gaussian random variable implies that the conditional variance decreases which explains why, in all our numerical examples with an indicator function, the highest eigenvalues ranked in $\ell$-order are simply the smallest eigenvalues. However, for multimodal failure functions, we may expect the conditional variance to increase and that the highest eigenvalues ranked in $\ell$-order are actually the largest ones. For multimodal problems, one may want to consider different parametric families of auxiliary densities, and so it would be interesting to see whether Theorem~\ref{thm1} can be extended to more general cases.
	
	\section*{Acknowledgement}
	The first author is currently enrolled in a PhD program co-funded by ISAE-SUPAERO and ONERA—The French Aerospace Lab. Their financial support is gratefully acknowledged. This work is part of the activities of ONERA - ISAE - ENAC joint research group.
	
	

	\bibliographystyle{siamplain}
	\bibliography{biblio}

\end{document}